\documentclass[
showpacs,
amsmath,amssymb,
aps,
pra,
twocolumn,
]{revtex4-1}

\usepackage{soul}
\usepackage{bibunits}
\usepackage{graphicx}
\usepackage{dcolumn}
\usepackage{bm}
\usepackage[breaklinks,colorlinks = true,linkcolor = red,urlcolor=blue,citecolor=red]{hyperref}
\usepackage{multirow}
\usepackage{array}
\usepackage{booktabs}
\usepackage{ctable}
\usepackage{upgreek}
\usepackage{epsfig}
\usepackage{mathrsfs}
\usepackage{amssymb}
\usepackage{amsbsy}
\usepackage{color}
\usepackage{cancel}
\usepackage{pifont}
\usepackage{marginnote}
\usepackage{float}
\usepackage{verbatim}
\usepackage{subfigure}
\usepackage{bbm, dsfont}
\usepackage{lineno}
\usepackage{amsmath}

\newcommand{\bcen}{\begin{center}}
\newcommand{\ecen}{\end{center}}
\newcommand{\btab}{\begin{tabular}}
\newcommand{\etab}{\end{tabular}}
\newcommand{\bdes}{\begin{description}}
\newcommand{\edes}{\end{description}}

\newcommand{\beq}{\begin{equation}}
\newcommand{\eeq}{\end{equation}}
\newcommand{\bea}{\begin{eqnarray}}
\newcommand{\eea}{\end{eqnarray}}

\newcommand{\bary}{\begin{array}}
	\newcommand{\eary}{\end{array}}
\newcommand{\benum}{\begin{enumerate}}
	\newcommand{\eenum}{\end{enumerate}}
\newcommand{\bitem}{\begin{itemize}}
	\newcommand{\eitem}{\end{itemize}}

\newcommand{\bK} { \mbox{\boldmath $K$}}

\newcommand{\bra}[1]{{\langle #1 |}}
\newcommand{\ket}[1]{| #1 \rangle}

\newcommand{\eqn}[1] {eqn.~(\ref{#1})}

\newcommand{\sect}[1] {Section~\ref{#1}}
\newcommand{\Sect}[1] {Section~\ref{#1}}

\newcommand{\Fig}[1]{Fig.~\ref{#1}}

\newcommand{\ci}{\mathbbm{i}}
\makeatletter

\newcommand{\Rmnum}[1]{\expandafter\@slowromancap\romannumeral #1@}
\makeatother

\newcommand{\signum}[0]{\mathop{\mathrm{sign}}}

\newcommand{\crt}[2]{{\st{#1}\color{blue}{#2}}}

\newcommand{\titlename}{
Statistics tuned entanglement of the boundary modes in coupled Su-Schrieffer-Heeger chains
}

\begin{document}
	
	\title{\titlename}
	
	\author{Saikat Santra}
	\email{saikat.santra@icts.res.in}
	\affiliation{International Centre for Theoretical Sciences, Tata Institute of Fundamental Research, Bengaluru 560089, India}
	\author{Adhip Agarwala}
	\email{adhip.agarwala@icts.res.in}
	\affiliation{International Centre for Theoretical Sciences, Tata Institute of Fundamental Research, Bengaluru 560089, India}
    \author{Subhro Bhattacharjee}
    \email{subhro@icts.res.in}
    \affiliation{International Centre for Theoretical Sciences, Tata Institute of Fundamental Research, Bengaluru 560089, India}
%


\date{\today}

\begin{abstract}

We show that mutual statistics between quantum particles can be tuned to generate emergent novel few particle quantum mechanics for the boundary modes of symmetry-protected topological phases of matter. As a concrete setting, we study a system of pseudofermions, defined as quantum particles with tunable algebra, which lie on two distinct Su-Schrieffer-Heeger (SSH) chains. We find that as the mutual statistics of the particles are tuned -- the boundary modes present in the two chains gets non-trivially entangled showing a sudden jump in their mutual entanglement entropy. We further show that, such tuning of statistics engenders a first-order transition between two topologically non-trivial phases which differ in the behavior of inter-chain entanglement. Using a combination of analytical and numerical techniques and effective modeling, we uncover the rich physics that this system hosts. The results are of particular relevance in context of the study of the effective low energy quantum mechanics of topological edge modes in one hand and their recent realization in ultracold atoms on the other. This then provides for controlled manipulation of such low energy modes.

 \end{abstract}

\maketitle

\section{Introduction}
Topological phases of matter show a multitude of exotic phenomena with crucial implications for the theoretical framework of understanding condensed matter phases on one hand, and material sciences with technological perspectives on the other\cite{Wen_RMP_2017,Ludwig_PS_2015, Chiu_RMP_2016,Hasan_RMP_2010, Qi_RMP_2011, Ando_JPSJ_2013,Yang_NatMat_2012,Beenakker_ARCMP_2013, Vergniory_arXiv_2018}. Characteristically, several such systems specifically in one spatial dimension, hosts symmetry protected topological(SPT) boundary {\it{zero}} modes\cite{Su_PRB_1980,Su_PRL_1979, Kitaev_PU_2001} as the manifestation of the non-trivial quantum entanglement in these systems. The low energy physics of these systems are then governed effectively by the properties of the few  boundary-modes. 

This {\it emergent} few-particle quantum mechanics of the novel boundary modes is extremely rich and forms the essential ingredient for estimating the usefulness as candidates for material realisation of quantum computing\cite{Kitaev_PU_2001,Alicea_RPP_2012,mourik2012signatures, Das_NatPhys_2012,Nayak_RMP_2008, Hyart_PRB_2013, Aasen_PRX_2016} where such boundary modes serve as {\it qubits}\cite{Boross_PRB_2019, Zaimi_arXiv_2019, Mei_PRA_2018}. Theoretical studies examining such boundary modes in coupled-wire systems\cite{Baeriswyl_PRB_1983, Padavi_PRB_2018, Zhang_PRA2_2017, Nersesyan_PRB_2020}, their tunability in junctions \cite{Harper_PRR_2019, Zaimi_arXiv_2019, Alicea_RPP_2012}  and attempts to entangle them non-trivially are been vigorously pursued in this regard\cite{Alicea_RPP_2012, Boross_PRB_2019, Zaimi_arXiv_2019, Longhi_PRB_2019, Lang_NJP_2017, Stanescu_JPCM_2013}. The complementary issue at the experimental front of controlled manipulation and tuning of such boundary modes are also being currently explored in wide variety of different materials \cite{Nadj_Science_2014, Rokhinson_NP_2012, Dumitrescu_PRB_2015, Deng_Science_2016, Albrecht_Nature_2016, mourik2012signatures, Deng_NL_2012, Liu_PRL_2012, Klinovaja_PRL_2013, Lutchyn_NRM_2018} as well as in cold atom systems \cite{Li_NatCom_2013, Meier_NatCom_2016, Leder_NatComm_2016, Xie_NQI__2019, Sylvian_Science_2019}.

It is crucial therefore to identify potential microscopic ``knobs" that can be ``tuned" to manipulate the low energy physics of these  topologically non-trivial boundary modes, and investigate their interplay with the symmetries. In addition to this tuning of few particle quantum mechanics, these knobs can further engineer novel phase transitions in the underlying many body system.

\begin{figure}
\centering
\includegraphics[width=1.0\linewidth]{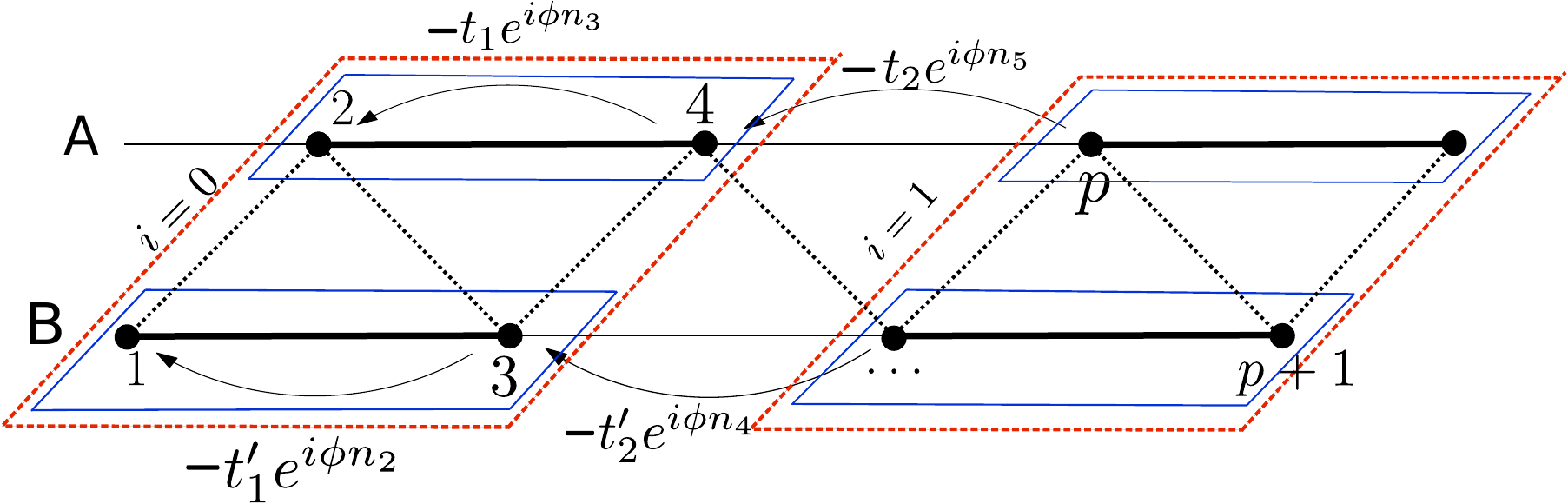}
\caption{{\bf Model:}  Schematic figure showing two SSH chains (A and B) labelled by site indices $p$. $i$ labels the four  
site unit cell where $t_1$ and $t_1'$ ($t_2$ and $t_2'$) are the intra (inter) unit cell hopping strengths as shown. Every hopping term has an additional phase $\phi$ that depends on the fermionic occupation of an intermediate site lying on the other chain -- this implements the pseudofermionic statistics (see text).}
\label{fig:schematic_SSH}
\end{figure}

In this paper, we investigate the above issues in context of the paradigmatic Su-Schrieffer-Heeger (SSH) model for polyacetylene \cite{Su_PRL_1979, Su_PRB_1980} that stabilises symmetry protected topological boundary modes. We examine an interesting tuning parameter that is of relevance in context of recent experimental \cite{Sylvian_Science_2019, Leinhard_PRX_2020} and theoretical \cite{Zuo_PRB_2018, Greschner_PRA_2018, Agarwala_PRB_2019}  developments -- the {\it generalised algebra} of the interacting fermions --  and study it in context of two coupled SSH chains as shown in  \Fig{fig:schematic_SSH}. The {\it statistical} tuning parameter, $\phi$ (see \eqn{eq_comm}), allows the degrees of freedom to smoothly transform from being fermionic to (hard core) bosonic and vice-versa in one spatial dimension.  Such degrees of freedom, referred as {\it pseudofermions} \cite{Agarwala_PRB_2019}, are generalizations of ``anyons" in one dimension \cite{Leinaas_NC_1977,Lieb_PR_1963,Kundu_PRL_1999, Pasquier_IMS_1994, Ha_PRL_1994,Ha_NPB_1995, Aneziris_IJMP_1991, Posske_PRB_2017,Frau_arXiv_1994,  Baz_IJMP_2003, Haldane_PRL_1991, Wu_PRL_1994,Murthy_PRL_1994,Batchelor_PRL_2006, Guo_PRA_2009, Eckholt_NJP_2009,Eckholt_PRA_2008, Greschner_PRL_2015,Forero_PRA_2016, Zhang_PRA_2017, Lange_PRL_2017,Lange_PRA_2017, Forero_PRA_2018, Zuo_PRB_2018}.  Quite remarkably, this anyonic physics has been recently realized experimentally in a cold atomic setting \cite{Leinhard_PRX_2020}. 

In our system, we find that as $\phi\in [0,\pi]$ is tuned, two topologically non-trivial SSH chains gets mutually entangled. In particular we uncover the rich low energy physics of the  many-body boundary modes of a  finite, but long, system -- as is relevant for the experiments. Crucially, $\phi$ allows modulation of the entanglement properties of the boundary modes residing on the two chains. The bulk, in the mean time, undergoes a first order transition at $\phi=\pi/2$, between the two topologically non-trivial phases.  We use a combination of numerical techniques (exact diagonalization (ED), density-matrix renormalization group (DMRG)) and analytical methods to uncover the physics of this system with particular emphasis to the emergent quantum mechanics of the boundary modes. 

The rest of this work is organised as follows. In \sect{sec:Model} we introduce the model and discuss its various relevant symmetries. In \Sect{manybodymodes} we discuss how this system stabilizes many-body boundary modes and study its dispersion as a function of $\phi$. In \Sect{sec:twop} we develop an effective theory for these boundary modes; in particular we study the case when both the chains contain one pseudofermion each and investigate how the boundary modes belonging to two chains gets non-trivially entangled showing a sudden jump in mutual entanglement. We further develop an understanding of this physics using few-particle quantum mechanics of effective boundary modes. We further examine the stability of these boundary modes to disorder and symmetry breaking perturbations.  In \Sect{sec:halffilled} we begin investigating the half-filled many body system.
Here we find that, while $\phi$ keeps the system topologically non-trivial -- there are in fact two distinctive phases at $\phi=0$ and $\phi=\pi$, which are separated by a first order phase transition.  Using many-body trial wavefunctions we identify the nature of the many-body ground states and further discuss the role of symmetry breaking perturbations on this many-body system.  In \Sect{sec:discussion} we summarize our results and provide concluding comments. In Appendices, we present additional results on DMRG calculations and role of disorder in these systems.   

\section{SSH Model for pseudofermions}
\label{sec:Model}

We start by introducing the {\it pseudofermions} \cite{Agarwala_PRB_2019} and their generalised algebra. Consider a two dimensional local Hilbert-space at each site, $p$, of a one dimensional lattice that are created and annihilated by second quantized {\it pseudofermions} operators $a_p^\dagger$ and $a_p$ respectively with $N_p=a_p^\dagger a_p$ being the number operator. The generalised algebra is now given by
\bea
&a_q a_p + a_p a_q e^{ \ci \phi~\signum(p-q)} =& 0 \nonumber\\
&a_q a^\dagger_p + a^\dagger_p a_q e^{- \ci \phi~\signum(p-q)}=& \delta_{pq}
\label{eq_comm}
\eea
where $\signum(0)=0$ gives an onsite fermionic algebra. The off-site algebra can be tuned from fermionic to (hard core) bosonic by tuning {\it statistical parameter} $\phi\in [0,\pi]$. Particles satisfying the above algebra have been previously dubbed as pseudofermions\cite{Agarwala_PRB_2019}-- a nomenclature that we continue to use in the present work.

We take two decoupled SSH chains of pseudofermions given by the Hamiltonian
\bea
H &=& -\sum_{i=0}^{L-1} \left[t_1 a^\dagger_{4i+2}a_{4i+4} + t_2 a^\dagger_{4i+4}a_{4i+6}+{\rm h.c.}\right] \notag\\ && - \sum_{i=0}^{L-1}\left[t'_1 a^\dagger_{4i+1}a_{4i+3} + t'_2 a^\dagger_{4i+3}a_{4i+5} + {\rm h.c.}\right]
\label{eqn:pseudoHam}
\eea
where $i={0,\ldots, L-1}$ sums over $L$ unit cells of four site each such that total number of sites (labelled by $p=1,\ldots,N$) $N=4L$ (see \Fig{fig:schematic_SSH}). $t_1 (t'_1)$ and $t_2 (t'_2)$ are the hopping amplitudes on the odd and even bonds of the upper $A$ (lower $B$) chain. We can consider all these parameters to be distinct, however we will restrict them shortly.

The $\phi=0$ limit clearly gives two decoupled free fermion chains whose properties are well known from the seminal works starting with those of Su, Schrieffer and Heeger~\cite{Su_PRB_1980, Su_PRB_1980}. Here we summarise the relevant part of these results for completeness. Choosing a two site unit cell for each of the chains separately (see \Fig{fig:schematic_SSH}) the Hamiltonian for chain $A$ (even indexed sites) and $B$ (odd indexed sites) can be straightforwardly diagonalized to obtain the single particle spectra for periodic boundary conditions :
\bea
E_A(k) &=& \pm \sqrt{ (t_1 + t_2 \cos(k))^2 +  (t_2 \sin(k))^2} \\
E_B(k) &=& \pm \sqrt{ (t'_1 + t'_2 \cos(k))^2 +  (t'_2 \sin(k))^2}
\eea
with $k\in[-\pi,\pi]$. At half-filling this system is gapped at all values of $t_1, t_2,t'_1$ and $t'_2$, except when $t_1=t_2 (t'_1=t'_2)$. In the gapped phase, the ground state hosts a topological band with a non-trivial winding number when $t_1<t_2$($t'_1<t'_2$) \cite{Pershoguba_PRB_2012}. In such a  topological phase, each individual chain under open boundary conditions, where the couplings between the $i=(L-1)^{th}$ unit cell and $i=0^{th}$ unit cell are removed, can host two {\it degenerate} single particle boundary modes which are localized at the two ends of the open chain.

The localization length of these boundary modes depends on the bulk gap and is given by $\zeta = 1/\ln \frac{t_2}{t_1} $ on chain A and similarly $ \zeta'= 1/\ln \frac{t'_2}{t'_1}$ on chain B. In any finite sized system, these localized states however hybridize leading to bonding and anti-bonding orbitals, exponentially close in energy by a gap scale  $ \sim \exp(-L/\zeta)$ and $ \sim \exp(-L/\zeta')$ for chains A and B respectively. At this point we find it convenient to restrict parameter space by choosing $t'_1= \gamma t_1$ and $t'_2 = \gamma t_2$ where we take $\gamma=2$ unless otherwise stated. This particular parameterisation proves helpful for energetic reasons as discussed below and does not affect the generality of our results. Importantly, given the recent realization of the SSH model in ultracold atoms\cite{Sylvian_Science_2019}, such parameterization opens up ways for controlled manipulation of the boundary modes.

Finite $\phi$ poses an interacting problem, and this becomes explicit, following Ref.~\onlinecite{Agarwala_PRB_2019}, once we recast \eqn{eqn:pseudoHam} in terms of spinless fermions created and annihilated by $c_p^\dagger$ and $c_p$ respectively through a fractional Jordan-Wigner transformations :
\begin{align}
	c_p=K_p a_p,\quad c_p^\dagger=a_p^\dagger K_p^\dagger~~{\rm with}~~K_p=e^{-\ci \phi \sum_{q < p} n_{q}}.
	\label{eq_fjw}
\end{align}
whence the Hamiltonian in \eqn{eqn:pseudoHam} becomes 
\beq
H= H_A + H_B
\label{eqn:Ham}
\eeq
where $H_A$ describes the hoppings on chain A, which now has $\phi$ dependent terms that depend on the site occupancies of chain B, i.e.~,
\bea
H_A = &-&\sum_{i=0}^{L-1} \left[t_1 e^{\ci\phi n_{4i+3}} c^\dagger_{4i+2}c_{4i+4} +  {\rm h.c.} \right] \notag \\ 
&-&\sum_{i=0}^{L-1} \left[ t_2 e^{\ci\phi n_{4i+5}} c^\dagger_{4i+4}c_{4i+6} + {\rm h.c.} \right] 
\eea
and similarly $H_B$ describes the hoppings on chain B with $\phi$ dependent terms that depend on the site occupancies of chain A, i.e.~,
\bea
H_B = &-&\sum_{i=0}^{L-1} \left[t'_1 e^{\ci\phi n_{4i+2}} c^\dagger_{4i+1}c_{4i+3} + {\rm h.c.} \right] \notag \\ 
&-&\sum_{i=0}^{L-1} \left[
t'_2 e^{\ci\phi n_{4i+4}} c^\dagger_{4i+3}c_{4i+5} + {\rm h.c.}\right].
\eea
Here, $c_p$ now obeys usual fermionic anti-commutation algebra. At any $\phi$, the number density  $n_p=c_p^\dagger c_p=N_p$ and hence the filling remains unchanged under the transformation (\eqn{eq_fjw}).

A distinct feature of \eqn{eqn:pseudoHam} and hence \eqn{eqn:Ham} is the fact that the total number of particles in each chain $A$ and $B$ remain independently conserved leading to a $U_A(1) \times U_B(1)$ symmetry for the system. This is in spite of the interaction between the $c$-fermions of the two chains mediated by $\phi\neq 0$ through the physics of correlated hopping \cite{Agarwala_PRB_2019}. This is reminiscent of the coulomb drag in bi-layer systems \cite{Narozhny_RMP_2016, Lee_PRL_2016, Li_PRL_2016}.

Continuing with the symmetries of the system, a single SSH chain made of $2L$ sites \cite{Sylvian_Science_2019} say of the form, 
\beq
H_S= -\sum_{i=0}^{L-1} \left[t_1 c^\dagger_{2i+1}c_{2i+2}  + t_2 c^\dagger_{2i+2} c_{2i+3} + {\rm  h.c.} \right]
\label{singleSSH}
\eeq
is symmetric under staggered charge conjugation operation implemented by an anti-unitary operator
\beq
{\cal C} = \left[ \prod_{i} \Big(c^\dagger_{2i+1} + c_{2i+1} \Big) \Big(c^\dagger_{2i+2} - c_{2i+2} \Big) \right] \circ \bK
\label{eqn:charge}
\eeq
(where $\bK$ is the complex conjugation operator) such that
\begin{align}
    {\cal C} c_{p} {\cal C}^{-1} \rightarrow\left\{\begin{array}{ll}
    c^{\dagger}_p & \forall~ p\in {\rm odd}\\
    -c^{\dagger}_p & \forall~ p\in {\rm even}\\
    \end{array}\right.
\end{align}

This can be generalised for the present case for the two chains and arbitrary $\phi$ by defining 

\bea
U_o &=& \left[ \prod_i \Big(c^\dagger_{4i+1} + c_{4i+1} \Big) \Big(c^\dagger_{4i+3} - c_{4i+3} \Big) \right] \notag \\
&& \times \left[ \prod_i e^{-\ci \frac{\phi}{2}\{(4i+2)n_{4i+2}+(4i+4)n_{4i+4}\}} \right] \circ \bK 
\label{symm_opt1}\\
U_e &=& \left[ \prod_i \Big(c^\dagger_{4i+2} + c_{4i+2} \Big)  \Big(c^\dagger_{4i+4} - c_{4i+4} \Big) \right] \notag \\ && \times  \left[ \prod_i e^{-\ci \frac{\phi}{2}\{(4i+1)n_{4i+1}+(4i+3)n_{4i+3}\}}\right] \circ \bK
\label{symm_opt2}
\eea

such that 
\begin{align}
    U_o c_{4i+a} U_o^{-1}\rightarrow\left\{\begin{array}{ll} c^{\dagger}_{4i+1} & \forall~i~{\rm and}~a=1\\
    e^{\ci \frac{\phi}{2}(4i+2)}c_{4i+2} & \forall~i~{\rm and}~a=2\\
    -c^{\dagger}_{4i+3} & \forall~i~{\rm and}~a=3\\ 
    e^{\ci \frac{\phi}{2}(4i+4)}c_{4i+4} & \forall~i~{\rm and}~a=4\\
    \end{array}\right.
\end{align}    
and 

\begin{align}
    U_e c_{4i+a} U_e^{-1}\rightarrow\left\{\begin{array}{ll}   e^{\ci \frac{\phi}{2}(4i+1)}c_{4i+1} & \forall~i~{\rm and}~a=1\\
 c^{\dagger}_{4i+2} & \forall~i~{\rm and}~a=2\\
    e^{\ci \frac{\phi}{2}(4i+3)}c_{4i+3}  & \forall~i~{\rm and}~a=3\\ 
-c^{\dagger}_{4i+4} & \forall~i~{\rm and}~a=4\\
    \end{array}\right.
\end{align}  

Note that while $U_o(U_e)$ leads to staggered charge-conjugation in the odd B (even A) chain, it leads to a multiplication by a site dependent phase in the even A (odd B) chain. Further, both $U_e$ and $U_o$ reduces to ${\cal C}$ operators for the respective chains $A$ and $B$ when $\phi=0$. 
Among other symmetries, it is useful to note that presence of non-zero $\phi$ breaks the time-reversal symmetry in the system. Also, the spectrum at any $\phi$ is symmetric about zero energy -- this is due to the bipartite structure of the Hamiltonian (\eqn{eqn:Ham}). A unitary symmetry operation of the form $c^{\dagger}_{4i+a} \rightarrow c^{\dagger}_{4i+a} (a=1,2)$, $c^{\dagger}_{4i+a} \rightarrow -c^{\dagger}_{4i+a} (a=3,4)$ takes $H \rightarrow -H$ and guarantees that $E$ and $-E$ states occur in pairs.  At $\phi=0$ the boundary modes of each of the SSH chains are protected by $\cal C$ leading to a total of four localised edge modes. We now discuss the fate of the system at finite $\phi$ which, as we shall show, leads to non-trivial entanglement of the boundary modes of the two chains.

\section{Interactions and and many-body boundary modes}
\label{manybodymodes}

Absence of time-reversal and presence of $U_o U_e$ places our system in class D of the Altland-Zirnbauer \cite{Altland_PRB_1997,Ludwig_PS_2015,Agarwala_AOP_2017} classification. However for general $\phi \neq 0$, the system is interacting and does not necessarily belong to one of the ten classes of free fermion SPTs\cite{Ryu_NJP_2010, Kitaev_AIP_2009}. Here we are mainly interested in the fate of the boundary modes of a finite but long system at general values of $\phi$. This allows us to engineer protocols for manipulating their properties using $\phi$ as a tuning parameter interpolating between the two limits of free fermions and hard-core bosons. Therefore we immediately focus on these boundary modes while the complete characterisation of the bulk physics is deferred to \sect{sec:halffilled}. 

In the topological phase ($t_1<t_2$, $\phi=0$), a half filled open system should comprise of many body boundary modes at low energies. To estimate the number of such states it is worthwhile to note that the free fermionic limit ($\phi=0$) hosts four single particle boundary modes close to zero energy which reside on the boundary while all other states ($4L-4$) reside in the bulk. Therefore, in a system with number of fermions $=2L-2$ ($L-1$ on each chain) only the single-particle bulk states would get occupied with no contribution from the boundary modes, generating a unique ground state. However, in a half-filled system i.e.~, with number of fermions $(=2L)$, two particles now occupy the single particle boundary modes. The latter can be achieved in four ways while maintaining the condition that each chain has one boundary mode occupied. These thus correspond to {\it four} many-body quasi-degenerate ground states for a half-filled system at $\phi=0$ and $t_1<t_2$ with open boundary conditions. When $t_1>t_2$ ({\it i.e~,} in the trivial regime) we have a unique ground state for both $2L-2$ and $=2L$ fermion number sectors, given the absence of any boundary modes.

\begin{figure}
	\centering
	\includegraphics[width=1.0\linewidth]{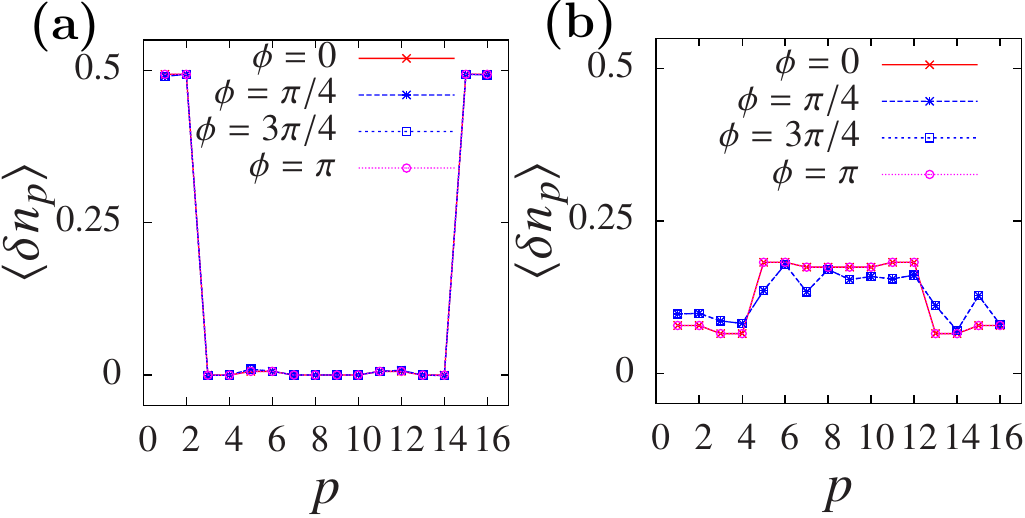}
	\caption{{\bf Boundary density:}  Difference of particle densities ($\langle \delta n_p \rangle$) between the ground states for $\tilde{N}_{pf}=2L$ and $\tilde{N}_{pf}=2L-2$ particle number sector, as a function of sites $p$ for various values of $\phi$ in (a) the topological regime ($t_1=0.1=1-t_2$) (b)  the trivial regime ($t_1=0.9=1-t_2$). In (a) the values are normalized over all the four quasi-degenerate ground states for $\tilde{N}_{pf}=2L$ sector (see text) (ED,$\gamma=2,N=16=4L$).}
	\label{fig:Den_extrct}
\end{figure}

The discussion above shows that for $t_1<t_2$ a difference in particle densities at any site between the (four) ground states for $2L$ particles and the ground state for $2L-2$ particles should reveal the boundary character of the many-body wavefunctions at half-filling, if any. In contrast, for trivial phase ($t_1>t_2$, $\phi=0$) given the absence of any single particle boundary modes, one expects such a density difference between the two ground states at $2L$ and $2L-2$ particle sectors wouldn't have any specific boundary character.  Extending this understanding at $\phi \neq 0$, we  calculate the difference in particle densities at every site for number of pseudofermions, $\tilde{N}_{pf}=2L$ system (averaged over the four lowest energy states for $t_1<t_2$ and unique state for $t_1>t_2$) and for $\tilde{N}_{pf}=2L-2$ system. We find that even when $\phi \neq 0$ and $t_1<t_2$ (topological phase) the residual densities (normalized over all the four degenerate states) $\sim$ 0.5 on the boundary sites (see \Fig{fig:Den_extrct} (a)); however when $t_1>t_2$ (trivial phase) the residual densities lie in the bulk showing that the system with $\tilde{N}_{pf}=2L-2$ should be considered as two delocalized holes which distribute uniformly over the bulk sites (see \Fig{fig:Den_extrct} (b)). {\it This shows that the boundary modes remain intact for the entire parameter regime of $\phi\in [0,\pi]$ and thereby continuously interpolating between the ``fermionic" ($\phi=0$) and the (hard-core) ``bosonic" ($\phi=\pi$) limits.}

\begin{figure}
	\centering
	\includegraphics[width=1.0\linewidth]{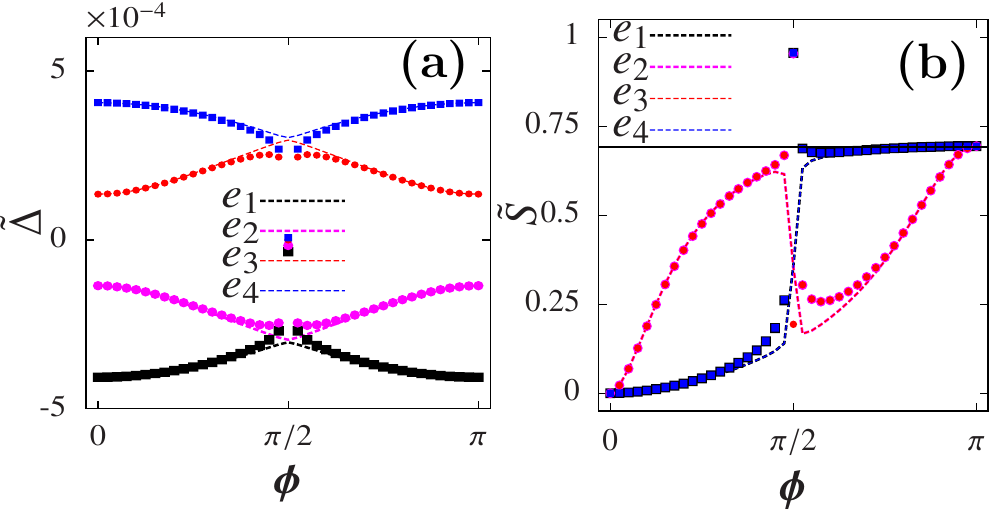}
    \caption{{\bf Many body boundary excitations:} 	(a) Energy of the four boundary modes for a half-filled system with open boundary conditions as measured by the {\it two-particle} gap defined by \eqn{tpg} (shown using points). (b) The corresponding mutual entanglement between the chains A and B with bulk contribution removed (see \eqn{tpgS}) (shown using points) saturates to a value of $\ln(2)$ (shown by a solid line) (ED,$N=4L=16,t_1=0.1,t_2=1-t_1,\gamma=2$.) Dashed lines ($e_1 \ldots e_4$) represent the results from two particle problem as discussed in \sect{sec:twop}. The erratic behavior of energies and entanglement entropy exactly at $\phi=\frac{\pi}{2}$ (in both Figs.~(a) and (b)) is due to a bulk  transition which we discuss in \sect{sec:halffilled}.}
	\label{fig:Extraction}
\end{figure}

Given the existence of such boundary modes for $t_1<t_2$ and $\phi \neq 0$ we track the energetics of the quasi-degenerate sub-space by calculating the symmetrized {\it two-particle} gap given by 
\begin{align}
\tilde{\Delta}=& E(\tilde{N}_{pf}=2L) \nonumber \\ &~~~~~ -\frac{1}{2}\Big (E(\tilde{N}_{pf}=2L+2) + E(\tilde{N}_{pf}=2L-2) \Big)
\label{tpg}
\end{align}
where $E$ is the ground state energy for $\tilde{N}_{pf}$ pseudofermions in a system of $L$ unit cells with open boundary conditions. As defined, $\tilde{\Delta}$ measures the energy cost to populate the boundary modes of our system in the topological phase (see \eqn{tpg}) at a given value of $\phi$.  
The resulting behavior of $\tilde{\Delta}$ is shown in \Fig{fig:Extraction}(a). The figure also shows that the modulation of energy of the boundary modes occurs at a much smaller scale compared to the single particle bulk gap (see \Fig{fig:figure4}(d)) and hence proving that the boundary modes retain their sanctity for all $\phi$ except at $\phi = \frac{\pi}{2}$. 

To further probe the nature of the four energetically isolated (from the bulk) boundary-modes at the ends of the chain, we calculate, for the half filled system, their {\it mutual quantum entanglement} defined as follows. We calculate the bipartite entanglement of chain A with respect to chain B \cite{Eisert_RMP_2010}, $S(\tilde{N}_{pf})$ for $\tilde{N}_{pf}=2L$ (half filled) as well as $\tilde{N}_{pf}=2L\pm 2$. In order to distill the contribution from the boundary modes, we then subtract any residual bulk contribution by defining
\begin{align}
\tilde{S}=& S(\tilde{N}_{pf}=2L)\nonumber\\
&~~~- \frac{1}{2}\Big ( S(\tilde{N}_{pf}=2L+2) + S(\tilde{N}_{pf}=2L-2)  \Big)
\label{tpgS}
\end{align}

The resulting behavior is shown in \Fig{fig:Extraction}(b).  While, as is expected, the modes are unentangled (between chains A and B) at $\phi=0$; each of these states however display a mutual entanglement of $\sim \ln(2)$ at $\phi=\pi$\crt{}{.} {\it This is the {\it central} result of this work -- where $\phi$, the parameter which tunes the algebra of pseudofermions,  can engineer an effective interaction between the subspace of many-body boundary modes leading to nontrivial entanglement between the two physically separated SSH chains.} The effective dynamics of these boundary modes, in the low energy subspace, leads to an emergent few body quantum mechanics which we now focus on.

\section{Effective theory of the boundary modes}
\label{sec:twop}

Having provided the evidence for the existence of the boundary modes for general $\phi$ in the half filled system for $t_1<t_2$, we now develop an effective theory of these many body boundary modes which are energetically separated from the extended bulk states. This effective theory as we show below is correctly captured by a two particle problem where each of the chain is populated by just one particle residing in the boundary modes. In \Fig{fig:Extraction}, we plot the energies of the two particle states which are close to zero (corresponding to boundary modes) and their mutual entanglement entropy along with the results of the half filled system. The close agreement for all $\phi\neq \pi/2$ indicates the one-to-one correspondence between the boundary modes of the two-particle system and the half filled one as detailed below. In Section \ref{sec:halffilled} we shall discuss the phase transition at $\phi=\pi/2$ in the half filled system which is of course a many-body bulk-effect. However, both the energies and entanglement entropies for the two-particle system shows extreme sensitivity to this transition as generically expected from the bulk-edge correspondence. Following our discussion near \Fig{fig:Den_extrct}, at $\phi=0$ and in the topological regime one expects four two-particle boundary states which are a direct product of single-particle boundary modes of each chain. In the next section we discuss the effect of non-zero $\phi$ on these two-particle states.

\begin{figure}
	\centering
	\includegraphics[width=0.925\linewidth]{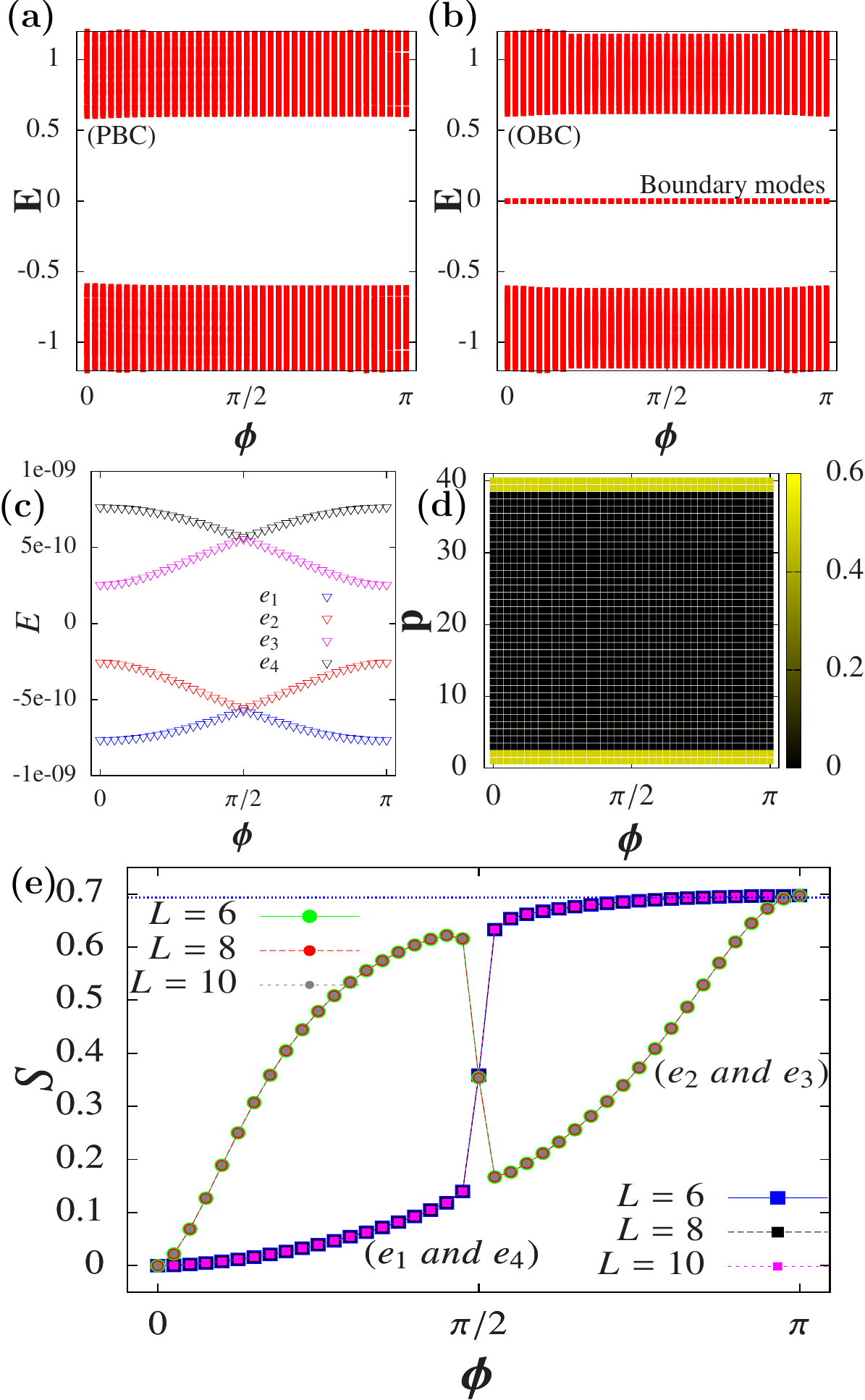}
	\caption{{\bf Two particle problem and boundary  modes:} Energies for a two particle system under (a) periodic boundary conditions (PBC) and (b) open boundary conditions (OBC) as a function of $\phi$.  (c) The near-zero energy states in (b) are zoomed-- showing four states (labelled $e_1-e_4$) which disperse as a function of $\phi$.  (d) The combined LDOS of the four states plotted as a function of position, $p$ show that the states close to $E=0$ are localised at the boundaries. (e) The mutual entanglement entropy ($S$) between the two chains as a function of $\phi$ for each of the four states. They reach a value of $\ln(2)$ at $\phi=\pi$ (shown by dashed line).	(ED,$t_1=0.1, t_2=0.9, \gamma=2$, for (a-d) $N=4L=40$)}
\label{fig:figure2}
\end{figure}

\subsection{Boundary modes and entanglement}

We numerically diagonalise the Hamiltonian in \eqn{eqn:Ham} for the above two particle set up to obtain their energy spectrum which is plotted as a function of $\phi$ for both periodic and open boundary conditions in \Fig{fig:figure2}(a) and (b) respectively. Unlike the periodic system, the system with open boundary conditions shows $4$ quasi-degenerate zero energy states, labelled ($e_1-e_4$), whose energies are weakly sensitive to $\phi$ as shown in \Fig{fig:figure2}(c). Our choice of $\gamma=2$ (defined above Eq. \ref{eq_fjw}) changes the relative bandwidths of the two chains in such a way that the two particle spectrum is gapped at $E=0$ for the periodic system, and existence of two particle boundary modes, if any, is clearly visible. 

In \Fig{fig:figure2}(d) we plot the LDOS for these four states showing that they indeed remain localised at the boundary. As remarked above, the boundary modes of the two-particle system very closely reproduces the excitation energies of the boundary of the half-filled system as was shown in \Fig{fig:Extraction}(a). We explore this one-to-one correspondence for the boundary modes to explore their properties in greater detail. It is worthwhile to note that these two-particle boundary modes are not the ground states of the two-particle problem and are separated in energy from all two particle bulk states by tuning $\gamma$. This is in contrast to the many-body boundary modes where such boundary modes become the quasi-degenerate ground state manifold for an half-filled system. Also given the presence of just one particle per chain the renormalization of the energies of the bulk two-particle states due to $\phi$ (see \Fig{fig:figure2}(a)) is not as drastic as the half-filled many-body states which we  discuss in \sect{sec:halffilled}.

The mutual entanglement between the boundary modes of chain $A$ and $B$ for the two particle system can be calculated in a straightforward manner as follows. We perform a singular value decomposition (SVD) of the two particle state, $|\Psi\rangle$, to obtain
\begin{equation}
\ket{\Psi} =\sum_{i,j=0}^{2L-1} a_{ij}\ket{2i+2}\ket{2j+1}  
\end{equation}
where $\ket{p}=c^\dagger_p\ket{0}$ is the single particle {\it fermionic} state  at position $p$ with the even and odd positions belonging to chain $A$ and $B$ respectively. The matrix ${\bf a}$ can be SVD diagonalized $\Lambda= U^{-1}{\bf a}V$ to obtain the diagonal eigenvalues $\Lambda_{i}$ \cite{Amico_RMP_2008}. The $\ket{\Psi}$ then can be expressed as
\begin{align}
\ket{\Psi} = \sum_{j} \Lambda_j \ket{\Psi_{jA}} \ket{\Psi_{jB}}
\label{eqn:1}
\end{align}
where $ \ket{\Psi_{jA}}=\sum_{i} U_{ij} |2i+2 \rangle$ and  $ \ket{\Psi_{jB}}=\sum_{i} (V^{-1})_{ji}  |2i+1 \rangle  $ form orthonormal vectors in single particle Hilbert space of both the chains separately. This decomposes the wavefunction in a way that the two sectors correspond to two distinct chains. $\Lambda_i$ characterizes the entanglement properties of this two particle state. In particular, a direct product state has $\Lambda_i = \delta_{i,0}$. More generally the entanglement entropy between two chains is given by \cite{Amico_RMP_2008}
\begin{align}
 S = -\sum_{j}\Big [| \Lambda_j |^2 \ln | \Lambda_j |^2 \Big ] \end{align}
 
In \Fig{fig:figure2}(e) we plot $S$ for the four boundary modes of the two-particle problem  ($e_{1},\ldots, e_4 < \Delta_{bw}$) (where  $\Delta_{bw}$ is the bandwidth of these modes as discussed below). Interestingly, while for all the states, the chains remain un-entangled at $\phi=0$, they saturate to a value of $\sim \ln 2$ at $\phi=\pi$ undergoing a jump at $\phi=\pi/2$. A systematic increase in $L$ doesn't change this functional dependence of $S$ on $\phi$, thereby reflecting that this result is indeed stable even at the thermodynamic limit (see \Fig{fig:figure2}(e)). Once again, the results of the two-particle mutual entanglement properties of the boundary modes are in one-to-one correspondence with that of the half filled case as shown in \Fig{fig:Extraction}(b). 

\begin{figure}
			\centering
			\includegraphics[width=0.9\linewidth]{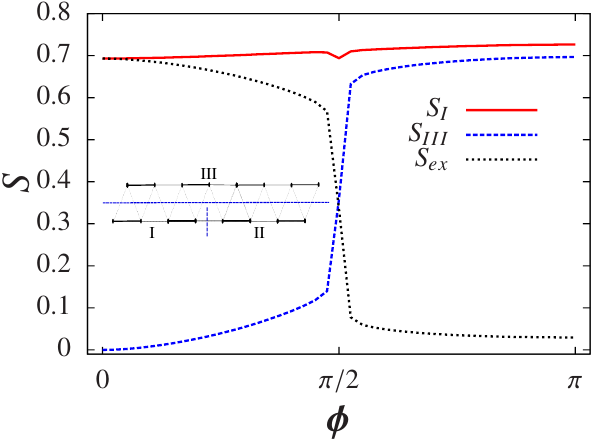}
			\caption{{\bf Intra-chain entanglement:} The behavior of entanglement entropy $S_I, S_{III}$ and $S_{ex}$ (see \eqn{excent}) (the regions I,II,III is shown in the inset) as a function of $\phi$ for the two particle system in the topological regime for the two particle state $e_1$ (see \Fig{fig:figure2}(c)).(ED$,t_1=0.1,t_2=0.9,\gamma=2,N=4L=16)$.}
			\label{fig:excl}
		\end{figure}

It is important to contrast the above  mutual (inter-chain) entanglement, with the intra-chain entanglement as a function of $\phi$. To this end, we partition the system into three regions (I, II, III) (see inset in \Fig{fig:excl}) with the bipartite entanglement between the particular part and the rest of the system being given by $S_I, S_{II}$ and $S_{III}$ respectively. While $S_{III}$ represents the mutual entanglement previously calculated (see \Fig{fig:figure2}(e)), $S_I$ represents the entanglement entropy of the left part of chain B with the rest of the system.  \Fig{fig:excl} shows that $S_I$ remains close to $\ln(2)$ as a function of $\phi$ indicating that the left boundary mode in $B$ continues to remain entangled throughout-- as is expected in a topological phase. We further calculate
\beq
	S_{ex}=S_{I}-S_{III}
	\label{excent}
	\eeq
which represents the exclusive entanglement entropy between the regions (I) and  (II). At $\phi=0$,  given the ground state is in a direct product state of two electrons residing in bonding orbitals in each of the chains, the entanglement entropy between left-half of chain A (B) and right half of chain A (B) is expected to be $\ln(2)$ which is confirmed in \Fig{fig:excl} for $\phi<\pi/2$. However, as seen in the figure, $S_{ex}$ jumps from $\sim \ln(2)$ to $\sim$zero at $\phi=\frac{\pi}{2}$ (see \Fig{fig:excl}). The almost perfect anti-correlation between $S_{ex}$ and $S_{III}$ shows that as the chains get mutually entangled, the exclusive entanglement between left and right parts of chain B(A) goes to zero in accordance with the monogamy of entanglement for the boundary modes akin to spin $S=\frac{1}{2}$ degree of freedom or spinless fermions.

\subsection{Minimal model for the boundary modes}

In order to develop possible protocols to manipulate the boundary modes for finite chains, it is desirable to develop an effective description only involving them that is operational at energy-scales much below the bulk gap. This takes the form of an effective few particle quantum mechanics with subtle features resulting from the underlying non-trivial nature of the bulk stemming from the underlying topological phase. In our case this leads to a four-site quantum mechanics whose features can be controlled by tuning $\phi$.

For a single finite SSH chain (see \eqn{singleSSH}) in its topological phase, there are two nearly degenerate eigenstates close to zero energy. These are the bonding ($+$) and the anti-bonding($-$) orbitals formed out of the linear combination of the boundary modes
\bea
c^\dagger_+ = \frac{1}{\sqrt{2}} \Big(c^\dagger_L +  c^\dagger_R \Big)~~~~~~c^\dagger_- = \frac{1}{\sqrt{2}} \Big(c^\dagger_L - c^\dagger_R \Big)
\label{left_right}
\eea
where $c^\dagger_L ( c^\dagger_R )$ creates an exponentially localized wavefunction on left (right) edge of the chain. $c_{+}$ and $c_{-}$ are energetically split by energy, \cite{Asboth_Notes_2016}
\beq
\alpha \approx \frac{t_1 t^2_2}{t^2_1+t^2_2} \times  e^{-(L-1)/\zeta},
\label{eq_alphaexp}
\eeq
where $\zeta = 1/\ln(t_2/t_1)$. Therefore, an effective Hamiltonian for {\it just} the boundary modes of a single SSH chain is $H_{\text{eff}} = -\alpha (c^\dagger_L c_R + {\rm h.c.})$.

\begin{figure}
	\centering
	\includegraphics[width=1.0\linewidth]{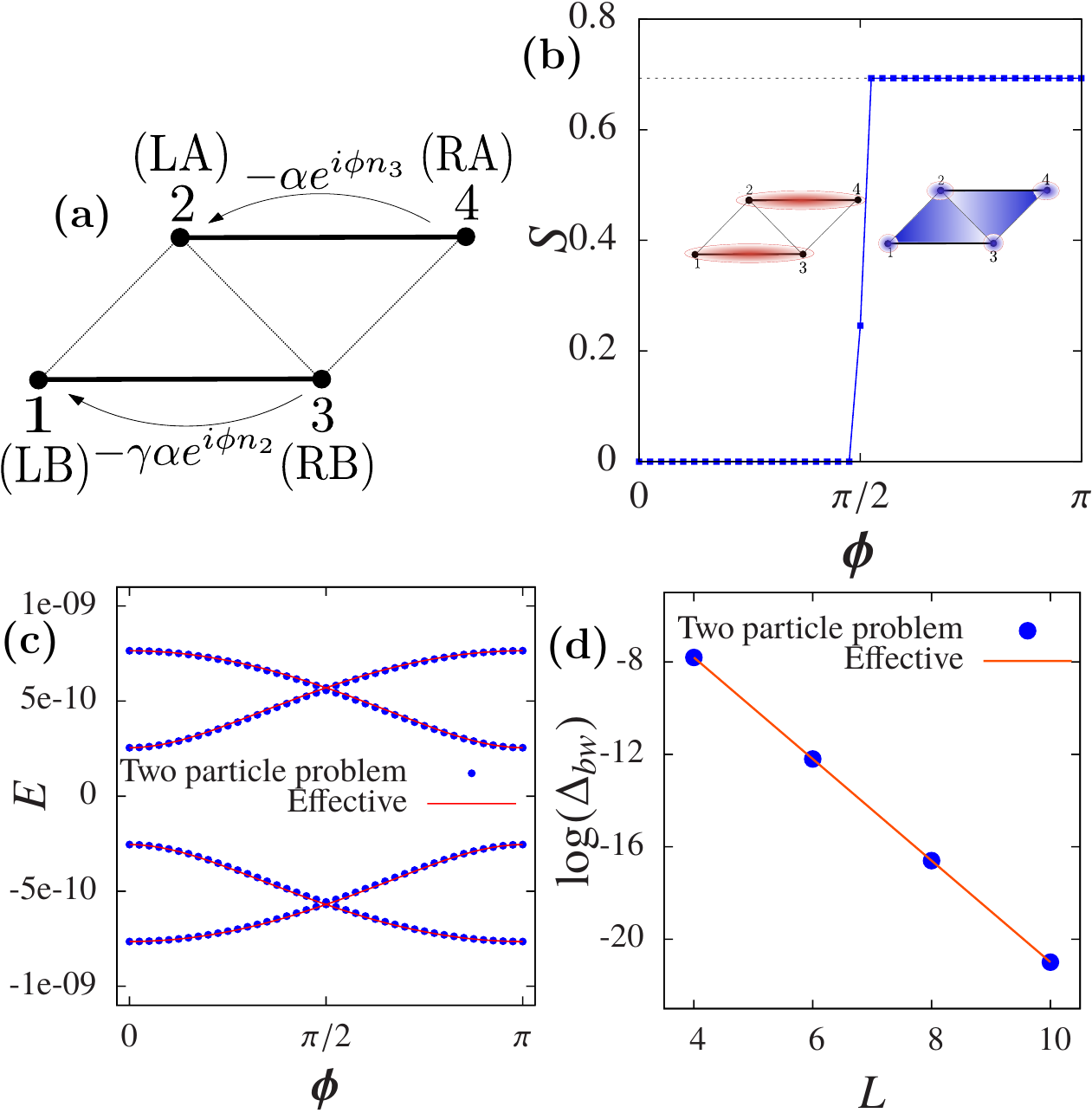}
	\caption{{\bf Minimal four-site model:} (a) Schematic of the 4-site model of the boundary modes. (b) Entanglement jump of $\ln(2)$ as a function of $\phi$ for the ground state of the four-site system (\eqn{eqn:effHam} with $\gamma=\alpha=1$) that changes from an un-entangled state (\eqn{exact1}, two red colored bonding orbitals) to an mutually entangled state (\eqn{exact2}, blue shaded plaquette). (c) For actual choice (see \eqn{eq_alphaexp}) of $\alpha, \gamma$ in \eqn{eqn:effHam} the spectra of this four site problem compared to the two particle boundary modes ($t_1=0.1,t_2=1-t_1,\gamma=2$, $N=4L=40$). (d) Behavior of the bandwidth of boundary modes ($\Delta_{bw}$) from the two particle problem and its comparison to that obtained from the effective Hamiltonian in \eqn{eqn:effHam}, as a function of system size $L$.}
	\label{fig:toymodel}
\end{figure}

For two chains: $A$ and $B$, we similarly have four single particle degrees of freedom  $\Big(c^\dagger_{L,(A/B)}, c^\dagger_{R,(A/B)} \Big)$ corresponding to left and right boundary modes on each chains A and B. Identifying  $c^\dagger_{LB},c^\dagger_{LA},c^\dagger_{RB},c^\dagger_{RA}$ with four 
``effective sites"  $\equiv (1,2,3,4)$ of a four site cluster (see \Fig{fig:toymodel}(a)), we form a two particle basis given by 
\bea
\ket { 1,  2} &\equiv& \ket{LB, LA} ~~~~~~~~ \ket {1,  4} \equiv \ket{LB, RA} \notag \\
\ket { 3,  2} &\equiv& \ket{RB, LA} ~~~~~~~~\ket { 3, 4} \equiv \ket{RB, RA} 
\label{twopbasis}
\eea
which reflects that both the chains have one particle each. 

In order to uncover the effective dynamics among these boundary modes, as tuned by $\phi$ dependent correlated hopping term, one can consider the case where a particle localized at right boundary of chain A ($RA$ - ``site 4") is brought to the left boundary ($LA$ -``site 2") via a hopping process $\propto t_1(t_1t_2)^{L-1}$. However, in the process the $\phi$ dependent term would contribute an overall phase which depends on the number density on the chain B up until it encounters the left boundary of chain B. This should lead to an effective term of the kind $\sim  c^\dagger_2 c_4 e^{i\phi n_3}$ since $n_3$ measures the density on the right boundary of chain B. Similarly considering the equivalent process on the other chain, the effective Hamiltonian at any $\phi$ is given by 
\beq
H_{\textbf{eff}} = - \alpha e^{\ci \phi n_{3}} c^\dagger_{2}c_{4} - \gamma \alpha e^{\ci \phi n_{2}} c^\dagger_{1}c_{3} + {\rm  h.c.}
\label{eqn:effHam}
\eeq
where $\gamma \alpha$ is the magnitude of the corresponding energy splitting between the boundary modes of the (finite) chain B ($t'_1=\gamma t_1, t'_2 = \gamma t_2$). The effective Hamiltonian, for $\phi=0$, reduces to
\begin{align}
H_{\text{eff}} (\phi=0) &=& -\alpha (c^\dagger_{LA} c_{RA} + {\rm h.c.})  -\gamma \alpha (c^\dagger_{LB} c_{RB} + {\rm h.c.}) 
\label{eqn:effRLLR}
\end{align}
which is trivially two copies of the edge Hamiltonian for two decoupled SSH chains, A and B. It is interesting to note that the four-site effective Hamiltonian of the boundary modes (see \eqn{eqn:effHam}) is also the Hamiltonian for a single unit-cell comprising of four sites in our parent Hamiltonian when $\alpha=t_1,t_2=0$  (see \eqn{eqn:Ham}).

The eigenspectrum of $H_{\textbf{eff}}$ and the mutual entanglement of the states (given a choice of $t_1, t_2, \gamma, L$) matches with the results for the full system almost exactly. The comparison of the eigenspectra for this effective four-site model and that of the boundary modes from the exact two-particle system is shown in \Fig{fig:toymodel}(c). Furthermore the fact that the dispersion of the boundary modes for this two particle system and its entanglement properties almost exactly matches the corresponding features of many-boundary modes (see \Fig{fig:Extraction}) shows that \eqn{eqn:effHam} is the effective Hamiltonian for the many-body boundary modes in this system as tuned by $\phi$. The discussion above also provides for the effective bandwidth of the two particle boundary states ($\Delta_{bw}$) given by $2|\alpha+ \gamma \alpha|$,
which, as expected, falls off exponentially with $L$ (see \Fig{fig:toymodel}(d)). Given the exponential fall in $\Delta_{bw}$ with increasing $L$, it is pertinent; especially in an experimental setting, to maintain $L$ and $\zeta$ length scales which can allow for tunability between the boundary modes even while resolving their individual energies. Apart from energetic scale $\alpha$ and the anisotropy factor $\gamma$ which denote the finite hybridization between the boundary modes of the finite chain, it is the presence of $\phi$ -- reflecting a  correlated hopping process -- that plays the central role in the effective boundary physics and entanglement characteristics. 

To understand this, we momentarily set $\alpha=\gamma=1$ in \eqn{eqn:effHam}. The resultant ground state is given by
\beq
\ket{\Psi_{1}} =\big(\frac{e^{\ci \frac{\phi}{2}}c^\dagger_{1}+c^\dagger_{3}}{\sqrt{2}}\big) \big(\frac{e^{\ci \frac{\phi}{2}} c^\dagger_{2}+c^\dagger_{4}}{\sqrt{2}}\big)\ket{\Omega_2}
\label{exact1}
\eeq
$\forall~\phi<\pi/2$ and
\beq
 \ket{\Psi_{2}}= \frac{1}{2} \big(- e^{\ci \phi}c^\dagger_{1} c^\dagger_{2} -\ci e^{\ci \phi/2} c^\dagger_{1} c^\dagger_{4}+\ci e^{\ci \phi/2}c^\dagger_{3} c^\dagger_{2}+c^\dagger_{3} c^\dagger_{4} \big)\ket{\Omega_2}
 \label{exact2}
 \eeq
 $\forall~\phi>\pi/2$ where $|\Omega_2\rangle$ describes the vacuum state of the effective quantum mechanics of the boundary modes. At $\phi=\pi/2$ the above two states are energetically degenerate and the corresponding energy-levels cross at $\phi=\pi/2$. In this case, for the ground state, up until $\phi=\pi/2$ the mutual entanglement between the two chains is identically zero and then jumps to $\ln(2)$ at $\phi=\frac{\pi}{2}$ (showed by dashed line in \Fig{fig:toymodel}(b)). This is expectedly so, given the form of $\ket{\Psi_{1}}$ (see \eqn{exact1}) is a direct product state between the two chains which can be shown schematically by two disjoint bonding-like orbitals on the two chains. This is to be contrasted with $\ket{\Psi_{2}}$ (see \eqn{exact2}) which is an entangled state (shown schematically in \Fig{fig:toymodel}(b)) where even when fermions are delocalized equally among the four sites, the state cannot be represented as a direct product (as shown by a blue shaded plaquette).

 Therefore we show that the effective low energy physics of the many body boundary modes (which in turn was captured by the physics of two-particle boundary modes; see \Fig{fig:Extraction}); is  actually that of a correlated hopping in a four-site cluster, albeit non-local. This allows for the entanglement tuning of the non-local modes that is facilitated by the non-trivial bulk (see \Fig{fig:Extraction}).
 
At this point we note that $|\Psi_1\rangle$ and $|\Psi_2\rangle$ are in fact eigenstates of the two symmetry operators $U_o$ and $U_e$ defined in \eqn{symm_opt1} and \eqn{symm_opt2} with the eigenvalues shown in the Table~\ref{symm_tab}. The level crossing at $\phi=\frac{\pi}{2}$, which leads to an entanglement jump, is therefore a reflection of the change in the symmetry representation of the ground state as $\phi$ is tuned above $\phi=\frac{\pi}{2}$. We shall discuss the implication of these ideas further in Section \ref{sec:halffilled} where we discuss the underlying phase transition in the many-body system.

 \begin{table}[h!]
	\begin{center}
		
		\label{tab:table2}
		\begin{tabular}{l|c|r} 
			\hline
			
			\textbf{}\vline & $\ket{\Psi_1}$&  \textbf{$\ket{\Psi_2}$} \vline \\
			
			\hline \hline 
			\vline $~$ $U_o$ & $e^{-\ci 5 \phi/2}$ & $\ci e^{-\ci 5 \phi/2}$ \vline \\
			\hline
			\vline $~$ $U_e$ & $e^{-\ci 2\phi}$& $- \ci e^{-\ci 2\phi}$
			\vline \\
			\hline
		\end{tabular}
		\caption{Eigenvalues of the symmetry operators $U_e$ and $U_o$ (\eqn{symm_opt1} and \eqn{symm_opt2}) for the two eigenstates $\ket{\Psi_1}$ and $\ket{\Psi_2}$ (see \eqn{exact1} and \eqn{exact2}).  $\ket{\Psi_1}$ ($\ket{\Psi_2}$) is the ground state for $\phi<\frac{\pi}{2}$ ($\phi>\frac{\pi}{2}$) for the effective Hamiltonian (see \eqn{eqn:effHam}) when $\alpha=\gamma=1$. }
		\label{symm_tab}
	\end{center}
\end{table}

We now briefly comment on a possible experimental protocol to measure the above mutual entanglement (Fig. \ref{fig:figure2}(e)) relevant to recent experiments\cite{Sylvian_Science_2019, Leinhard_PRX_2020} and especially applicable in this two particle setting. Following Ref.~\onlinecite{Abanin_PRL_2012}, the setup comprises of engineering a quantum switch with two states denoted by $\ket{\uparrow}, \ket{\downarrow} $ coupled to the Hamiltonian (see \eqn{eqn:Ham}). Now consider two values of the statistical parameter $\phi$ and $\phi'$ such that the complete system (SSH chains and the switch) is prepared in the ground states 
$\ket{\uparrow} \otimes \ket{\Psi (\phi)}$ and $\ket{\downarrow} \otimes \ket{\Psi (\phi')}$ where $\ket{\Psi (\phi)}$ and $\ket{\Psi (\phi')}$ are the ground states for $H(\phi)$ and $H(\phi')$ respectively. A system in $\ket{\Psi (\phi)}$  can be made to oscillate via Rabi oscillations to a state $\ket{\Psi (\phi')} $ using a tunneling term of the form $H' = \Gamma  (\ket{\uparrow} \bra{ \downarrow} + \ket{\downarrow}\bra{\uparrow})$ on the quantum switch. This leads to a characteristic Rabi frequency $\Omega$ where $\Omega  = \frac{\Gamma'   }{\hbar}$ and $\Gamma' = \Gamma  \langle \Psi (\phi) |\Psi (\phi') \rangle $\cite{Abanin_PRL_2012} which can then be experimentally measured. Note that this quantity is dependent on the overlap $\chi(\phi,\phi')= \langle \Psi (\phi) |\Psi (\phi') \rangle $. 

In order to track a mutually entangled state, it is particularly useful to discuss two choices of $\phi$ and $\phi'$: (a) $\phi=0$ and $\phi'=\pi$ where the $\chi(0,\pi)=\frac{1}{2}$ (calculated using the forms of wavefunctions shown in \eqn{exact1} and \eqn{exact2})  and (b) $\phi=\epsilon(\sim 0)$ and $\phi'= -\epsilon$  where $\chi \sim 1$. Note that the ground state for each of these values of $\phi$ and $\phi'$ have the same energy. However, the measurement of Rabi oscillations in (a) is half of that measured in (b). Unlike case (b), in case (a), $|\Psi(\phi')\rangle$ is an mutually entangled state. This characteristic halving of $\Omega$ would therefore signal generation of non-trivial {\it  mutual entanglement} between chains A and B.  While this result is expected for the four site cluster, and for the boundary modes for the two particle problem -- in a many body problem $\chi(\phi,\phi')$ falls exponentially with increasing system size (as we have checked numerically using many-body wavefunctions) rendering such a protocol potentially challenging to implement in an experimental setting.

\begin{figure}
	\centering
	\includegraphics[width=1.0\linewidth]{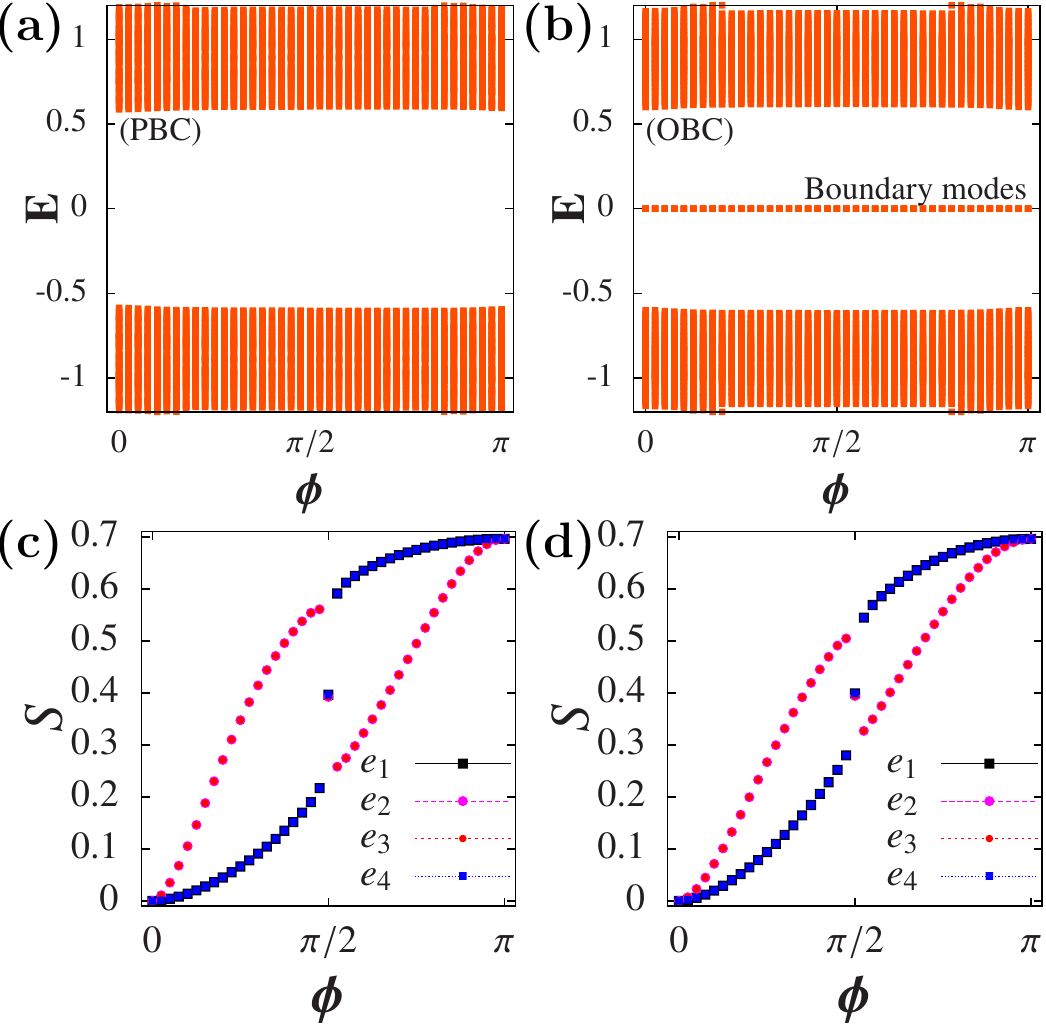}
	\caption{{\bf Stability to hopping disorder:} (a,b) The energy spectra for a two particle problem as a function of $\phi$ with hopping disorder strength $W=t_1/5$ under (a) periodic boundary conditions (PBC) and (b) under open boundary conditions (OBC). (c,d) The jump in mutual entanglement $S$ as a function of $\phi$ for the four two particle boundary modes ($e_1, \ldots, e_4$) for disorder strength $W=t_1/10$ in (c) and $W=t_1/5$ in (d). (ED, $t_1=0.1, t_2=0.9, \gamma=2, N=4L=40$)}	
	\label{fig:figure3}
\end{figure}

\subsection{Stability to disorder and other perturbations}
\label{sec:twoppert}

While our discussion until this point describes the existence of boundary modes and their tunability with respect to $\phi$, their topological origin should impart them stability against symmetry preserving disorder and other perturbations. Here we examine these effects in turns starting with the disorder.

For hopping disorder, the hopping amplitudes i.e. various $t$-s in \eqn{eqn:Ham} are modified to $\tilde{t}=t+\delta t$  where a small random number ($\delta t$) is uniformly drawn from a box distribution between $[-W,W]$ independently for each bond. We find that such a disorder does not destroy the boundary modes and their entanglement properties  (see \Fig{fig:figure3}) for the two particle problem. We have also checked that this property holds even for the many-body boundary modes (not shown). In contrast to site disorder\cite{Anderson_PR_1958}, hopping disorder retains the generalised charge-conjugating symmetries $U_e$ and $U_o$ separately. However, such protection of boundary modes is guaranteed only  until the point when the disorder scale is weak compared to the bulk gap scale in the system whence it drives a bulk phase transition to a trivially localised phase. In that limit both topological features and corresponding entanglement entropy changes significantly (also seen in other systems such as \cite{Laflorencie_PRB_2005, Pastur_JSP_2018}) (see Appendix~\ref{appenB} for a detailed discussion). 

\begin{figure}
	\centering
	\includegraphics[width=1.0\linewidth]{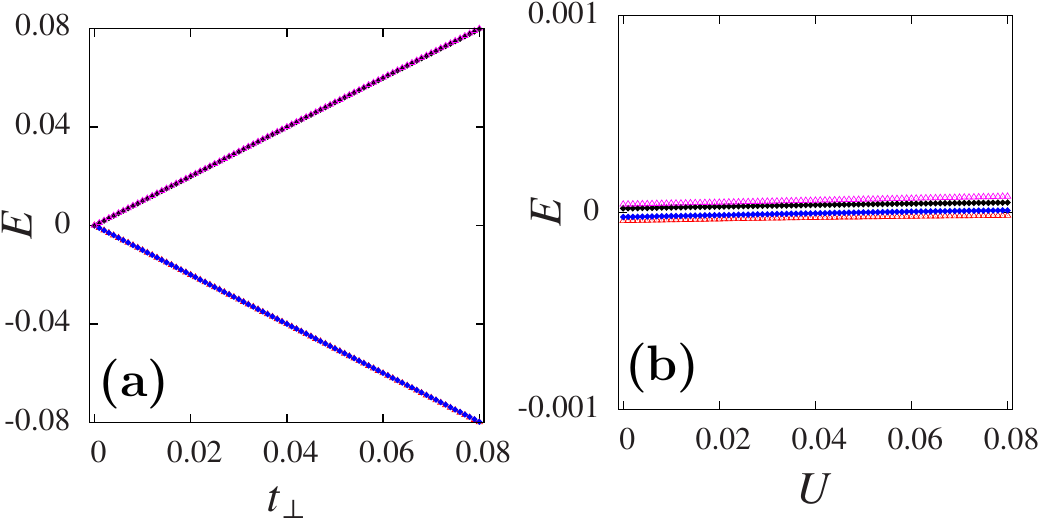}
	\caption{ {\bf Symmetry breaking/preserving perturbations:} (a) Energies of the single particle boundary modes as a function of inter-chain hopping strength $t_{\perp}$ at $\phi=0$  (see \eqn{eqn:tperp}). (b) Energies of boundary modes in presence of two particles, as a function of Hubbard interaction strength $U$ (see \eqn{hubbard}) at $\phi=\pi/4$. Here, an overall constant proportional to $U$ is subtracted.  (ED, $t_1=0.1,t_2=1-t_1,\gamma=2,N=4L=20$) }
	\label{fig:EnergySplit}
\end{figure}

We now turn to the role of {\it translationally invariant} symmetry preserving and symmetry breaking perturbations to \eqn{eqn:Ham} . Specifically, we study the effect of Hubbard interaction given by
\bea
H_U= \frac{U}{2} \sum_{i=0}^{L-1} \Big [  (\tilde{n}_{4i+1} + \tilde{n}_{4i+2}+ \tilde{n}_{4i+3} + \tilde{n}_{4i+4})^2-1 \Big] \notag \\+
U \sum_{i=0}^{L-1} \Big [ (\tilde{n}_{4i+4} + \tilde{n}_{4i+3}) (\tilde{n}_{4(i+1)+1} + \tilde{n}_{4(i+1)+2}) \Big] \notag \\
\label{hubbard}
\eea
where $\tilde{n}_p= n_p - \frac{1}{2}$ and the interaction term coupling the $i=(L-1)$ and the $i=0$ unit cell is kept such that the perturbing term doesn't break the symmetry of the parent periodic system. While this term is invariant under a product of $U_o$ and $U_e$, it is not invariant under them separately. This term further maintains the $U_A(1) \times U_B(1)$ number conservation symmetry of the system. Upto an overall chemical potential renormalization we find that the boundary modes do not split and remain quasi degenerate (exponentially close in system size $\sim e^{-L/\zeta}$). This can be contrasted with the case when the system is perturbed by an inter-chain hopping term of the kind,
\bea
	H_{t_{\perp}}= -\sum_{i=0}^{L-1} t_{\perp} \Big (  c^\dagger _{4i+1} c_{4i+2}+c^\dagger _{4i+2} c_{4i+3} \notag \\
	+c^\dagger _{4i+3} c_{4i+4} +c^\dagger _{4i+4} c_{4(i+1)+1}+ {\rm h.c.} \Big )
	\label{eqn:tperp}
\eea
Here $t_{\perp}$ breaks $U_o, U_e, U_o U_e$ and $U_A(1) \times U_B(1)$ symmetries. We find that such a perturbation immediately splits the boundary modes (see \Fig{fig:EnergySplit}(a)).

The two distinct behaviors for the two kinds of perturbations can be understood from the effective Hamiltonian for the boundary modes as discussed in \eqn{eqn:effRLLR} and \eqn{eqn:effHam}. $H_{t_{\perp}}$ effectively generates a hopping term of the kind
\beq
	H_{t_{\perp}}= - t_{\perp} \Big (  c^\dagger _{LB} c_{LA}+c^\dagger _{RB} c_{RA}+ {\rm h.c.} \Big )
	\label{eqn:interchain}
\eeq
immediately hybridizing the boundary modes, while $H_U$ generates a term of the kind 
\beq
H_U=U \big (\tilde{n}_{LB} \tilde{n}_{LA}+\tilde{n}_{LB} \tilde{n}_{RA}+\tilde{n}_{RB} \tilde{n}_{LA}+\tilde{n}_{RB} \tilde{n}_{RA} \big ) 
\label{eqn:hubbard}
\eeq
which acts as an identity term in \eqn{eqn:effHam}. Given the two distinctive behaviors of these perturbing terms, this analysis shows that the symmetry protecting this topological phase is rather a product of two anti-unitary operators $U_o U_e$, in combination with the number conservation $U_A(1) \times U_B(1)$. 

This completes our discussion of the effective quantum mechanics of the boundary modes. Our discussion of the two particle problem shows existence of quasi-degenerate eigen-modes in an open system when $t_1<t_2$ which reside on the boundaries and are robust to disorder and symmetry preserving perturbations; taken together, these results provide a comprehensive understanding that these modes are indeed of topological origins at any value of $\phi$.
In this section we further discussed their effective dynamics in presence of $\phi$, their entanglement properties and potential measurement protocols.

Having discussed this interplay of statistics, entanglement and symmetries -- on the boundary modes of this topological phase, we now turn to the many-body problem and discuss the bulk physics of the half filled system and understand its properties as a function of $\phi$.

\section{Many body system}
\label{sec:halffilled}

 The trivial and topological phase of a single SSH chain (see for e.g.,~\eqn{singleSSH}) is distinguished by the value of polarization (defined modulo $2$)
\beq
P = \frac{1}{\pi}\text{Im} \Big[ \ln \Big( \langle  \exp \Big( \ci \frac{2\pi}{L}\sum_{i=0,\ldots,L-1}^{j=1,2} x_{i} n_{2i+j} \Big) \rangle \Big) \Big]
\eeq
where $i$ runs over the unit cell index and $j$ over the sites within a unit cell  and $x_{i}$  describes the position of the $i^{th}$ unit cell \cite{Smith_PRB_1993, Resta_book_2007, Nakamura_PRB_2008, Watanabe_PRX_2018}. Also $\langle\cdots\rangle$ denotes expectation is taken over the many-body ground state of the system. The resulting value of polarization can be related to the geometric phase of the single-particle bands in an non-interacting system\cite{Resta_book_2007}. Given, presence of a non-trivial $\phi$ engineers interactions which doesn't allow the description of the many body state in terms of single Slater determinant; here, while one can not evaluate a single particle geometric phase (and corresponding winding number) -- but a many-body polarization can still be calculated for the half filled system as follows. 

\begin{figure}
	\centering
	\includegraphics[width=1.0\linewidth]{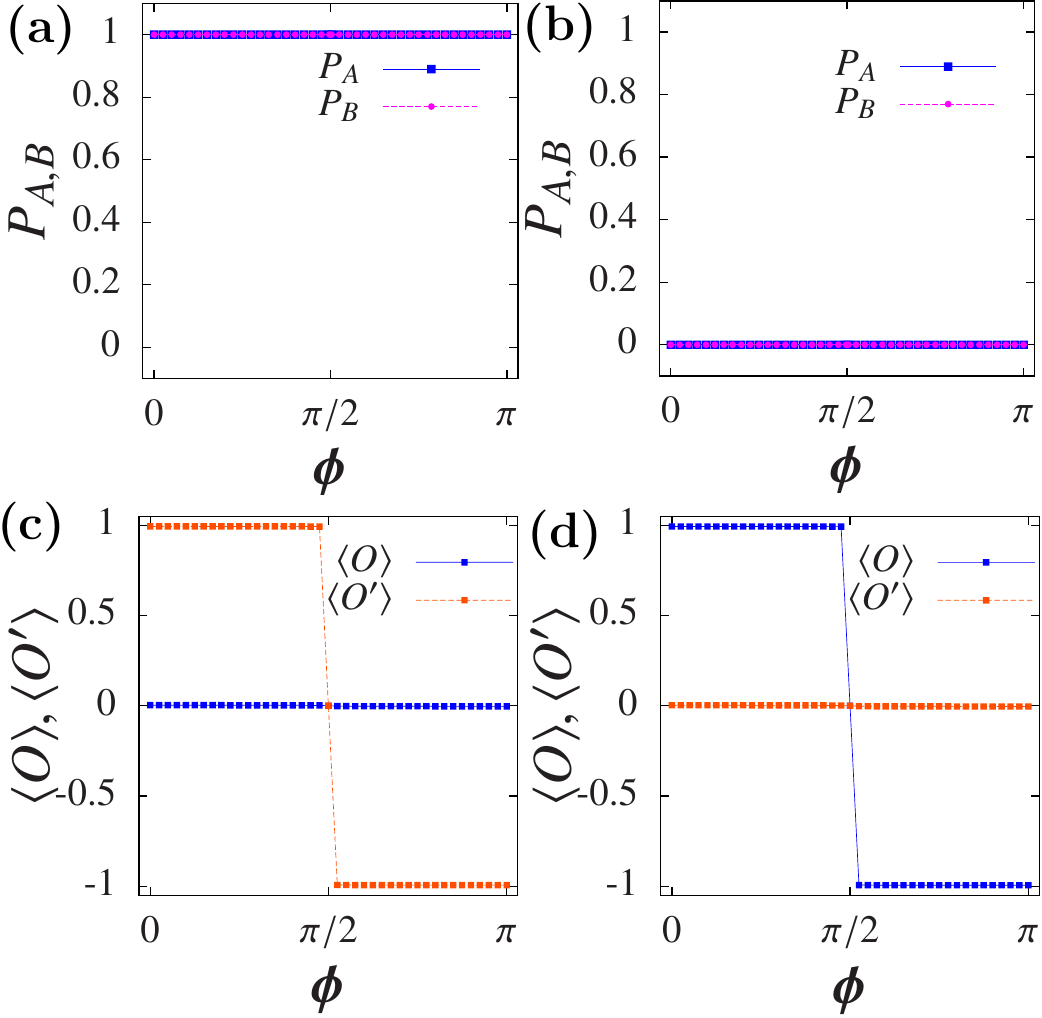}
	\caption{{\bf Polarization:} The behavior of $P_A$ and $P_B$ (see \eqn{eqn:pola}) with respect to $\phi$ for (a) $t_1=0.1$ (topological phase) (b) $t_1=0.9$ (trivial phase) (see text). The behavior of $\langle O\rangle, \langle O' \rangle$ (see \eqn{eqn:oopop}) as a function of $\phi$  in (c) $t_1=0.1$ and (d) $t_1=0.9$. (ED, $t_2=1-t_1,\gamma=1,N =4L=16$)}
	\label{fig:figure5}
\end{figure}

For our two-chain system, we calculate the polarization for each chain separately ($P_A$ and $P_B$) given by 
\bea
P_A &=& \frac{1}{\pi}\text{Im} \Big[ \ln \Big( \langle \exp \Big(\ci \frac{2\pi}{L}\sum_{j={2,4}}^{i=0 \ldots L-1} x_i n_{4i+j} \Big) \rangle \Big) \Big] \notag \\
P_B &=& \frac{1}{\pi}\text{Im} \Big[ \ln \Big( \langle \exp \Big(\ci \frac{2\pi}{L}\sum_{j ={1,3}}^{i=0 \ldots L-1}  x_i n_{4i+j} \Big) \rangle \Big) \Big].
\label{eqn:pola}
\eea
 We find that the values when evaluated over the ground state (defined modulo 2) continues to be non-trivially $1$ as a function of $\phi$ when $t_1<t_2$ (see \Fig{fig:figure5} (a)). It is not well defined at $\phi=\frac{\pi}{2}$ owing to a level crossing of the ground states which we shall shortly discuss. Our results therefore show that the many body topological phase which exists at $\phi=0$ is indeed stable all the way up to $\phi=\pi$ remaining independent of $\phi$.

Even as polarization remains non-trivial, is there an operator which differentiates the bulk phase $\phi <\frac{\pi}{2}$ and $\phi >\frac{\pi}{2}$? To construct such operators we draw insights from our study of the four site problem which was introduced as an effective problem for the boundary modes in \eqn{eqn:effHam}. While there these four ``sites" represented the effective boundary modes (see \Fig{fig:toymodel}); this effective Hamiltonian is also identical to the terms which appear in the Hamiltonian of a single unit cell (comprising of four sites) in our correlated fermionic Hamiltonian (see \eqn{eqn:Ham}). Specially at $t_2=0$ limit, our complete system can be understood as a direct product of four site clusters, each of which satisfies the Hamiltonian \eqn{eqn:effHam} with $\alpha=t_1$. Motivated by this connection we construct the following Hermitian operators, both of which commute with $U_o$ and $U_e$:

 {\small
 
\bea
	O &=& \frac{1}{L} \sum_{i}  \Big[ e^{\ci\frac{\phi}{2}}  c^\dagger_{4i+1} c_{4i+3} + {\rm h.c.} \Big]
	\Big[ e^{\ci\frac{\phi}{2}} c^\dagger_{4i+2} c_{4i+4} + {\rm  h.c.}\Big] \notag \\ 
	O'&=& \frac{1}{L} \sum_{i}  \Big[ e^{\ci\frac{\phi}{2}}  c^\dagger_{4i+3} c_{4i+5} + {\rm h.c.} \Big]
	\Big[ e^{\ci\frac{\phi}{2}} c^\dagger_{4i+4} c_{4i+6} + {\rm  h.c.}\Big].  \notag \\
	\label{eqn:oopop}
\eea}

$\langle O \rangle$ for a single four site cluster has an value $+1 (-1) $ for ground state wavefunction $|\Psi_1\rangle (|\Psi_2\rangle)$ (see \eqn{exact1} and \eqn{exact2}) and $O'$ is displaced by half a unit-cell from $O$. While at $\phi=0$, $O$ is a product of two hopping operators on the two bonds of chains A-B, at $\phi=\pi$ they appear as product of two current operators between the two chains. For a many body system, these effectively measures the location and the phase relationship of fermions on the bonds of the two chains respectively. Indeed for $\phi=0$, $\langle O \rangle =0~( \langle O' \rangle =1)$ in the topological phase ($t_1 <t_2$) and other way around in the non-topological phase ($t_1>t_2$).   Interestingly {\it even in the topological (trivial) phase, as $\phi$ is tuned, the sign of $\langle O' \rangle ~(\langle O \rangle)$ changes at $\phi =\frac{\pi}{2}$}\crt{}{.} (see \Fig{fig:figure5} (c) and (d)). This signals the non-trivial phase relationship of the bonding orbitals which is tuned by $\phi$ and suddenly changes its character at $\phi=\frac{\pi}{2}$.

\subsection{Entanglement and Gap}

\begin{figure}
	\centering
	\includegraphics[width=1.0\linewidth]{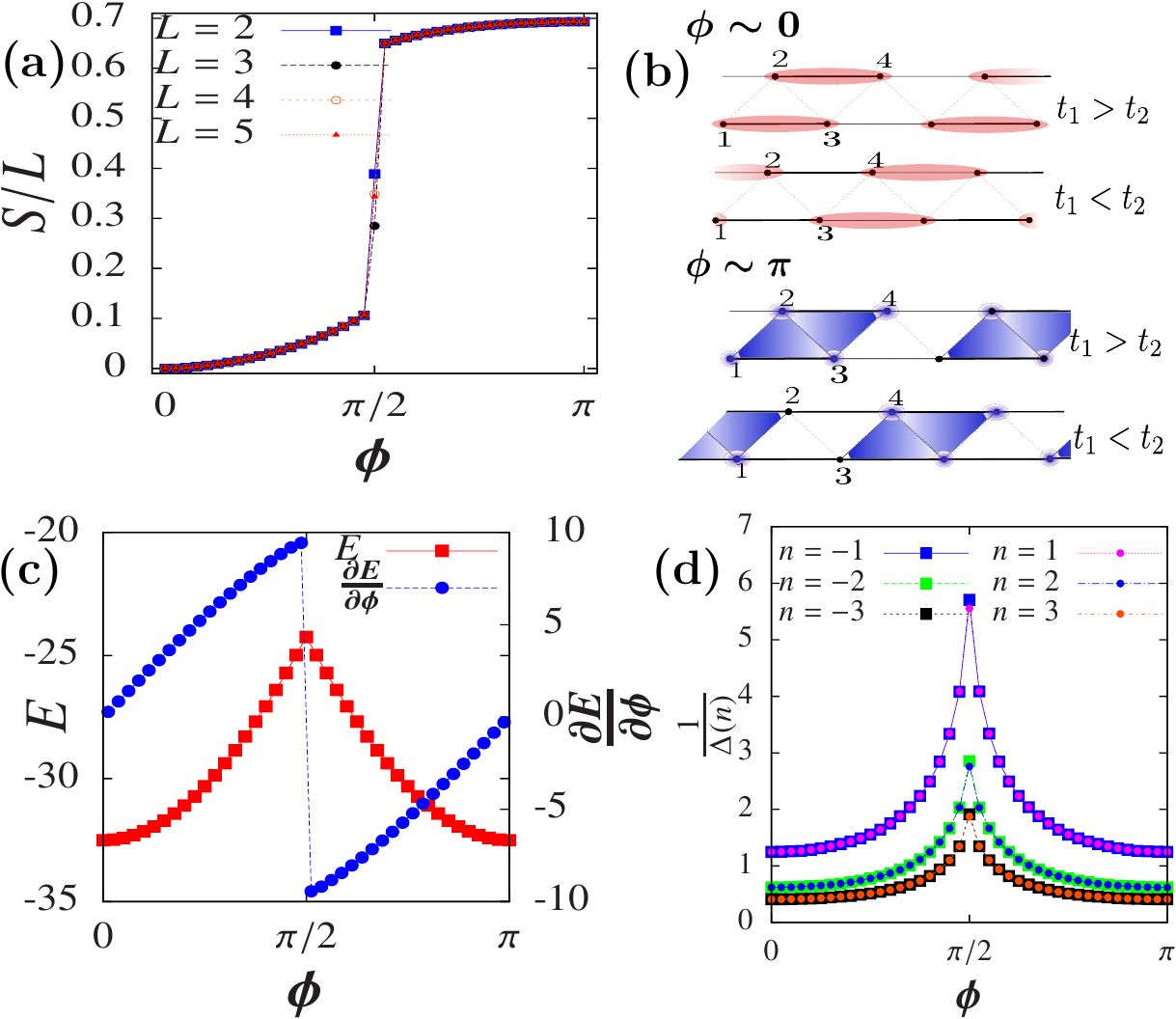}
	\caption{{\bf Half-filled periodic system:} (a) The behavior of mutual entanglement entropy $S$ as a function of $\phi$ shows a jump of $L \ln(2)$ at $\phi=\frac{\pi}{2}$ for different system sizes. (b) Schematic showing that the ground state for $t_1 \ll t_2$ and $t_1 \gg t_2$ can be interpreted as weakly coupled four site clusters  of unentangled states at $\phi \sim 0$ and  entangled states at $\phi \sim \pi$. (c)  Ground state energy $E$ and its derivative with respect to $\phi$ ($\frac{\partial E}{\partial \phi}$) as a function of $\phi$.  (d) Behavior of $1/\Delta(n)$ (see \eqn{spg}) as a function of $\phi$. (For (a) ED studies and for (c) and (d) $N=4L=48$, DMRG calculations). ($t_1=0.1, t_2=0.9,\gamma=2$, PBC) }
	\label{fig:figure4}
\end{figure}

In a system with open boundary conditions we had found that $\phi$ engineered a jump in the mutual entanglement between the boundary modes belonging to the two chains. In a periodic system, the ground state is unique (except at $\phi=\frac{\pi}{2}$) and we now investigate the behavior of mutual entanglement between the two chains in this ground state as a function of $\phi$. We find that such mutual entanglement jump continues at $\phi=\frac{\pi}{2}$, but with a value of $L \ln(2)$ which is {\it extensive} in system size (see \Fig{fig:figure4}(a)) given $L$ is the number of four-site unit-cells. This suggests that this entanglement is engineered through bulk states -- where each unit cell contributes a value of $\ln(2)$. This reflects what we had found as the properties of a single four-site cluster as discussed near \eqn{eqn:effHam}. The transition for the many-body state thus provides a natural understanding of the state, where deep in both the topological regime ($t_1\ll t_2$) and trivial regime ($t_1 \gg t_2$), the state can be interpreted as weakly coupled four-site clusters each of which contributing a $\ln(2)$ entropy (see \Fig{fig:toymodel}(a) and (b)) to the many body state once $\phi>\frac{\pi}{2}$ (see  \Fig{fig:figure4}(b)). This also provides the understanding for the effective four-site problem. This entanglement jump, as we have discussed above, is not specific to the topological phase; for instance even between the two trivial phases, $\phi$ can engineer an entanglement jump. Furthermore, while characterization of entanglement entropy depends on the microscopic degree of freedom used (for instance fermions or hard-core bosons), once defined with a chosen basis, its value and variation with $\phi$ is physical.

The sudden jump in entanglement at $\phi=\frac{\pi}{2}$ hints at an underlying thermodynamic phase transition. This is further confirmed by the behavior of the ground state energy as a function of $\phi$ and find that its first order derivative shows a distinctive jump at $\phi=\frac{\pi}{2}$, reflecting that the transition is of {\it first order} (see \Fig{fig:figure4}(c)). Interestingly even though system undergoes this first order transition the single particle gap $\Delta(n)$
defined as energy required to add or remove $n$ particles from the half-filled sector 
\beq
\Delta(n)= E(\tilde{N}_{pf}=2L+n)- E(\tilde{N}_{pf}=2L)
\label{spg}
\eeq
where $E$ is the ground state energy for $\tilde{N}_{pf}$ particles in a system of $L$ unit cells, remains finite as a function of $\phi$ across the tuning range (see \Fig{fig:figure4}(d)). That these gaps remain finite even as the system size is systematically increased (see Appendix~\ref{appenA} for details) shows that even when $\phi$ entangles the boundary modes and the bulk remains gapped to single particle excitations, the ground state undergoes a level crossing.

\subsection{Trial many-body wavefunctions}

Motivated by the structure of the wave-functions in the free SSH limit as well as that of the effective boundary Hamiltonian (Eq. (\ref{eqn:effHam})) we propose the following trial wave-functions for the many-body ground state for the Hamiltonian in eq. (\ref{eqn:Ham}) with $t_1=\gamma=1,t_2=0$. 

\beq
\ket{\Psi^{(L)}_1}=\prod_{i=0}^{L-1} \big(\frac{ e^{\ci \frac{\phi}{2}}c^\dagger_{4i+1}+c^\dagger_{4i+3}}{\sqrt{2}}\big) \big(\frac{e^{\ci \frac{\phi}{2}} c^\dagger_{4i+2}+c^\dagger_{4i+4}}{\sqrt{2}}\big)\ket{\Omega}
\label{eqn:var1}
\eeq
for $\phi<\pi/2$ and 
{\small 
	\bea
	\ket{\Psi^{(L)}_2}=\prod_{i=0}^{L-1}\big[ \frac{1}{2} \big(- e^{\ci \phi} c^\dagger_{4i+1} c^\dagger_{4i+2}  -\ci e^{\ci \frac{\phi}{2}} c^\dagger_{4i+1} c^\dagger_{4i+4} \notag \\ 
	+ \ci e^{\ci \frac{\phi}{2}}c^\dagger_{4i+3} c^\dagger_{4i+2}+c^\dagger_{4i+3} c^\dagger_{4i+4} \big)\big]\ket{\Omega}
	\label{eqn:var2}
	\eea}
for $\phi>\pi/2$. Here $|\Omega\rangle$ is the many-body vacuum.

\begin{figure}
	\centering
	\includegraphics[width=1.0\linewidth]{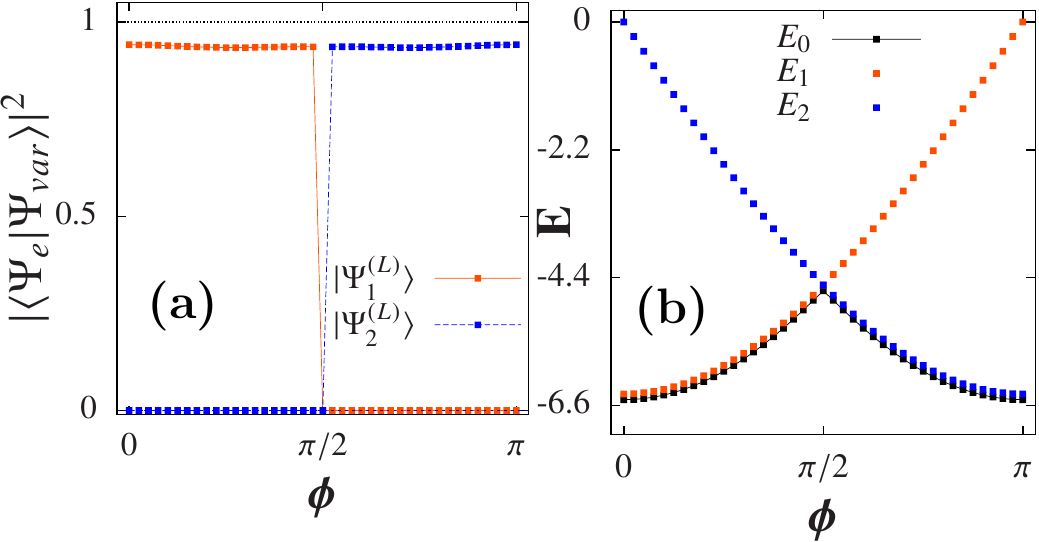}
	\caption{{\bf Trial wavefunctions:} (a) The behavior of the overlap ($\equiv |\langle \Psi_{e} |\Psi_{var}\rangle|^2 $) for the two choices of trial wave functions $|\Psi_{var}\rangle = |\Psi^{(L)}_{1}\rangle, |\Psi^{(L)}_{2}\rangle $ (see \eqn{eqn:var1} and \eqn{eqn:var2})  with the exact ground state wave function $|\Psi_{e}\rangle$, as a function of $\phi$. (b) Comparison of the corresponding energies for the the trial wavefunctions ($E_1 = \langle \Psi^{(L)}_1 |H| \Psi^{(L)}_1 \rangle$, $E_2 = \langle \Psi^{(L)}_2 |H| \Psi^{(L)}_2 \rangle$) and the exact ground state energy ($E= \langle \Psi_e |H| \Psi_e \rangle$) as a function of $\phi$. (ED, $t_1=0.8=1-t_2,\gamma=1,N=4L=16$)}
	\label{fig:figure65}
\end{figure}

The overlap with the exact ground-state for the half-filled system (even when $t_1\ne 1.0$), as obtained from ED, with $|\Psi^{(L)}_1\rangle$ ($|\Psi^{(L)}_2\rangle$) for $0\leq\phi<\pi/2$ ($\pi/2<\phi\leq\pi$) is notably high as shown in \Fig{fig:figure65} (a). The corresponding comparison of energies are shown in \Fig{fig:figure65} (b). This reflects that the ground state wave function is indeed well approximated by weakly coupled four site cluster wave functions which are un-entangled between chains A and B at $\phi=0$. However as $\phi$ is tuned, these un-entangled chains become entangled all throughout the bulk at $\phi=\frac{\pi}{2}$ undergoing the first-order transition. It is interesting to note that at $\phi=0$ and $t_1=1$, one has an extremely large degeneracy at many body zero energy, given each four site cluster has energy eigenvalues given by $-2,0,0,2$.  The many body spectrum can take at least ${L}\choose{\frac{L}{2}}$ (assuming $L$ is even) number of zero energy many body states. However $\phi$ chooses a particular combination of entangled states to engineer the many-body state which then forms the ground state when $\phi > \frac{\pi}{2}$. Therefore it is this change of ground state wave-function that is reflected both in the first-order transition, the jump in entanglement and corresponding jump in the value of the operators $O,O'$ (see \Fig{fig:figure5}). Unlike a finite temperature first order transition, which occurs with a latent heat corresponding to jump in the thermodynamic entropy; this quantum `first' order transition occurs with a jump in the mutual entanglement entropy.

\subsection{Breaking of $U_A(1)\times U_B(1)$ symmetry}
\label{sec:breakU1}

In the last section our ED, DMRG and studies using trial wavefunctions convincingly point to a first order transition which can be engineered using $\phi$. Importantly, in a system with open boundary conditions $\phi$ allows one to tune  the entanglement properties of the many-body boundary modes (\Fig{fig:Extraction}). The crucial symmetry protecting this physics has been $U_o U_e$ (see \eqn{symm_opt1} and \eqn{symm_opt2}) and the number conservation $U_A(1) \times U_B(1)$ on each chain as was discussed in \sect{sec:twop}(c).  In \sect{sec:twop}(c) we had looked at the effect of symmetry breaking and symmetry preserving perturbations on the two-particle problem and found that boundary modes would split in presence of an interchain hopping of strength $t_\perp$ (see \Fig{fig:EnergySplit}).In particular for this interchain hopping, the splitting is $\propto t_\perp$ which is a much larger scale compared to the finite size splitting scale at $t_\perp=0$ which is exponential small in system size. However, the actual mixing depends on the strength of perturbations with respect to the bulk gap and the boundary modes may survive for practical purposes for small perturbations, albeit now split. This occurs especially when the perturbing energy scales and probe fields are much smaller than the bulk gap. In this section we investigate such features and potential signatures which would be experimentally measurable even in presence of weak inter-chain hopping -- particularly when, such terms may be realistically present in any experimental setting \cite{Sylvian_Science_2019}.

\begin{figure}
	\centering
	\includegraphics[width=1.0\linewidth]{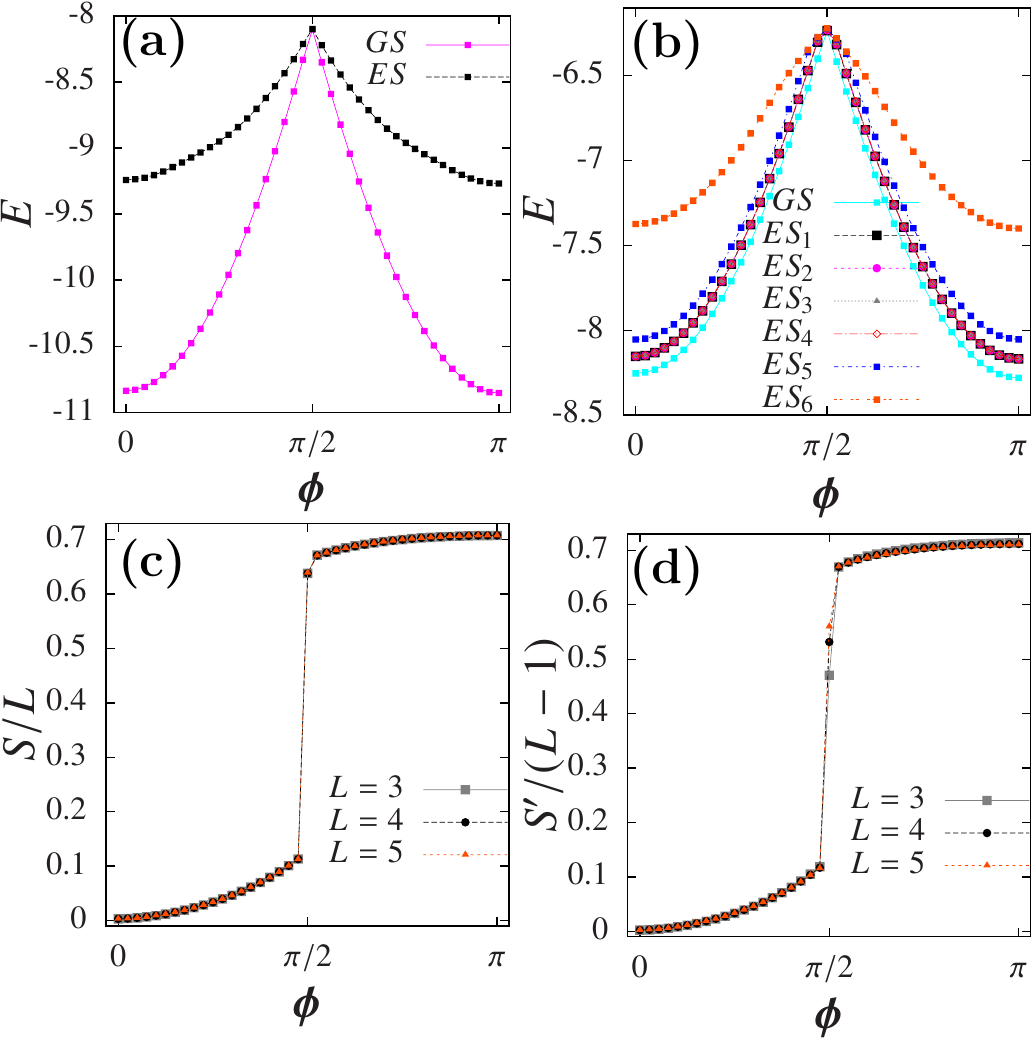}
	\caption{{\bf Breaking $U_A(1)\times U_B(1)$:} (a) Energies of the ground state and first excited state ($GS$ and $ES$) as a function of $\phi$ for a half-filled system under periodic boundary conditions (PBC) in presence of interchain hopping $t_\perp$ (see \eqn{eqn:tperp}) (b) Energies of the seven low lying energy states ($GS$ and $ES_1-ES_6$) for the same system but under open boundary conditions (OBC). (c) Behavior of entanglement entropy $S$ as a function of $\phi$ for the ground state under PBC for various system sizes. (d) Behavior of bulk entanglement entropy $S'=S-2\ln 2$ as a function of $\phi$ for various system sizes. (ED, $t_1=0.1, t_2=0.9, t_{\perp}=t_1/2, \gamma=2$, for (a,b) $N=4L=16$.)}
	\label{fig:figure6}
\end{figure}

In order to study this we add a quadratic hopping term $t_\perp$ (see \eqn{eqn:tperp}) along dashed lines  (see \Fig{fig:schematic_SSH}) to the parent Hamiltonian (\eqn{eqn:Ham}) and perform ED studies on the many-body problem. Given the number of particles on each chain are no longer conserved, simple counting in the free fermionic limit shows that one would expect {\it six} low energy states in an open system (6 ways of occupying boundary modes) when both the chains are in topological regime (instead of four as discussed before) and expect a unique ground state with a finite gap to excitations in a periodic system. This expectation is borne out in the ED studies (see \Fig{fig:figure6}(a) and (b)) where the many-body energy spectra for both periodic and open system is shown. In particular, we find that the low energy manifold in the open system is not exactly degenerate, but split $(\propto t_{\perp})$ due to a finite $t_\perp$ (consistent with results in \sect{sec:twop}(c)). However, these lowest six eigenstates still remains separated in energy to bulk excitations with an energy scale $t_1-t_2$ that characterises the bulk gap scale. This gap scale is characterized by the energy of the transition from ground state to sixth excited state, which remains finite (see \Fig{fig:figure6}(b)), except when near $\phi=\frac{\pi}{2}$ where the first order phase transition occurs.

We now investigate if the entanglement properties of the many-body states remain stable to inter-chain hopping. For a half-filled system under periodic boundary conditions and $t_1<t_2$ (topological regime) the system has a $L$ scaling for the entanglement jump (see \Fig{fig:figure6} (c)) as was the case when  $t_\perp=0$ (see \Fig{fig:figure4} (a)). Thus for a periodic system, presence of a finite $t_\perp$ does not change the jump in the mutual entanglement. However, under open boundary conditions, the many body ground state shows a $2\ln(2)$ entropy even at $\phi=0$ and then jumps to a value of $(L+1) \ln(2)$ at $\phi>\pi/2$. This behavior seems to have an extra component of $2\ln(2)$ entropy added to the expected behavior of  $(L-1)$ bulk unit cells which would contribute a entropy jump of $(L-1) \ln 2$ at $\phi=\frac{\pi}{2}$ (apart from the boundary sites the system has $L-1$ four site clusters, see \Fig{fig:figure4}(b)). 

This extra entanglement can be understood from the fact that any inter-chain coupling immediately hybridizes the left (right) boundary mode of chain A with left (right) boundary mode of chain B with an energy $t_\perp$. This leads to a formation of two bonding like orbitals at the ends of our zigzag ladder (see \Fig{fig:schematic_SSH}) coupling the two chains A and B dominating over the physics of effective correlated hopping process whose energetics is exponentially small in system size ($\propto \alpha$ (see \eqn{eqn:effHam})). These boundary bonding orbitals adds a $2\ln(2)$ contribution of mutual entanglement entropy to the bulk part independent of $\phi$. Thus, separating the bulk entanglement entropy part $\equiv S'=S-2 \ln(2)$  we can recover the expected scaling nature that is $\propto (L-1) \ln 2$ (see \Fig{fig:figure6} (d)).

Our discussion here, shows that -- even though $t_\perp$ splits the boundary modes; perturbatively, the physics of the transition, entanglement jump between two phases and existence of the low energy boundary sector under open boundary conditions remains stable when $t_\perp$ is smaller than the bulk gap.  We have further checked that many-body polarization remains non-trivial in presence of $t_\perp$ (not shown).

\section{Summary and Outlook}
\label{sec:discussion}

We now summarise our results. In this paper, using a combination of numerical and analytical calculations, we have shown that mutual statistics between quantum particles can be potentially ``tuned" to engineer novel quantum mechanics for the low energy sub-spaces formed out of topologically protected boundary modes in SPTs. We achieve, in particular, non-trivial tuning of the entanglement between the two one-dimensional chains which host topological phases using a statistical parameter $\phi$. Our  study therefore brings out an interesting interplay between quantum statistics, topological phases of matter, entanglement and symmetries.

\begin{figure}
	\centering
	\includegraphics[width=1.0\linewidth]{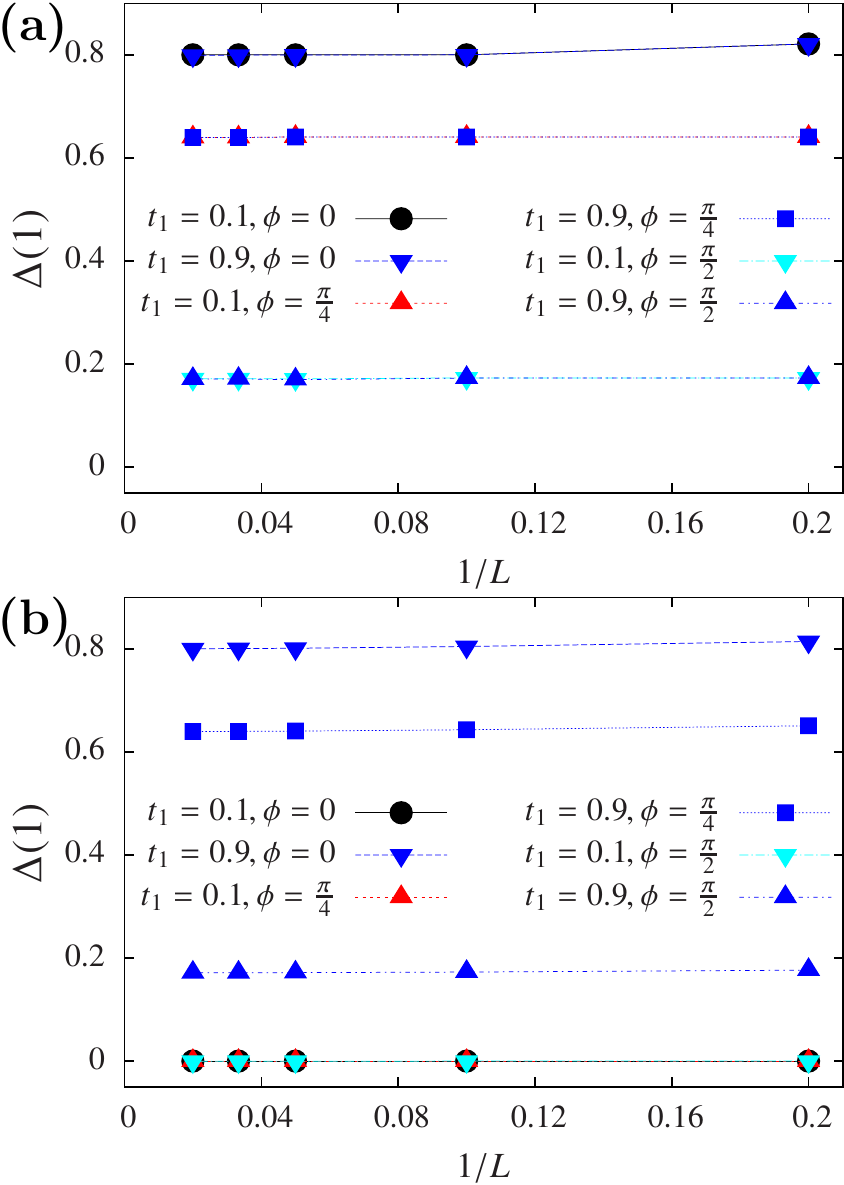}
	\caption{ {\bf Gap scaling:} The scaling of energy gap to add one particle ($\Delta(1)$, see \eqn{spg}) over the half-filled state with system size $L$ for (a) periodic and (b) open boundary conditions for various values of $t_1, t_2=1-t_1, \gamma=2$ and $\phi$. (DMRG results)}
	\label{fig:app1}
\end{figure}

As a concrete setting to achieve this end, we studied a system of  pseudofermions on two distinct SSH chains. We noticed that for a half-filled system the boundary modes gets mutually entangled showing a jump in the entanglement entropy of $\ln(2)$ at $\phi=\frac{\pi}{2}$ (see \Fig{fig:Extraction}). In \sect{sec:twop}, our study of just two particles in this system, provides a consistent understanding and the effective low energy quantum mechanics of the boundary phenomena. Moving on to the bulk physics then, we found that $\phi$ engenders a first-order transition between two topological phases, again with a corresponding jump in the mutual entanglement entropy between the bulk sites (see \Fig{fig:figure4}). While such a transition is not specific to topological phases, here, it allows to entangle the boundary modes non-trivially. We further investigated the role of disorder and various symmetry breaking and preserving perturbations to characterise the phase and its stability. 
Our work therefore points to an interesting class of phase and phase transitions, with potential technological implications, that can be engineered by tuning the algebra of these one-dimensional anyons.

\begin{figure}
	\centering
	\includegraphics[width=1.0\linewidth]{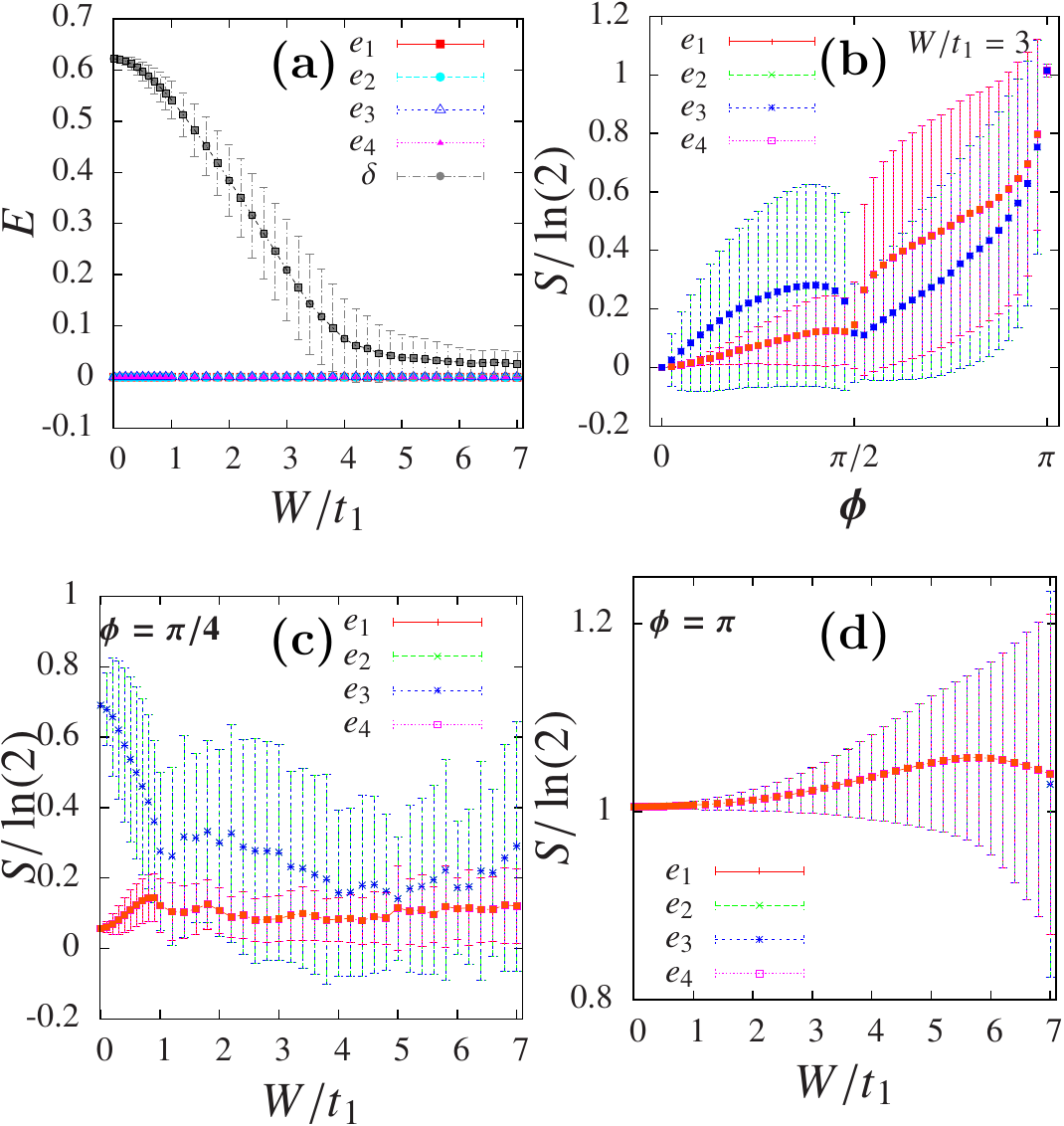}
	\caption{ {\bf Strong disorder:} (a) Behavior of average energies of the four close to zero energy states (labelled ($e_1-e_4$)) and the average energy gap to the bulk excited states defined as $\delta$ as a function of $W/t_1$ at $\phi=0$. The averaging is done over 50 samples. The error bars reflect the standard deviation. (b) Entanglement entropies of the boundary modes as a function of  $\phi$ for $W/t_1=3$. (c) and (d): The entanglement entropies of the boundary modes as a function of $W/t_1$ in (c) $\phi=\pi/4$ and (d) $\phi=\pi$. (ED, $N=4L=40,t_1=0.1=1-t_2,\gamma=2$)}
	\label{fig:app2}
\end{figure}

The possibility to tune the boundary modes and entangle them using statistical phase $\phi$ is particularly noteworthy given the exciting developments in the experimental forefront where the SSH model has been recently realized in a cold atomic setting -- \cite{Sylvian_Science_2019} using Rydberg atoms\cite{Browaeys_nature_2020}. This experimental setting has shown unprecedented control in populating individual boundary modes and their possible manipulation and measurement \cite{Elben_Science_2020}. The crucial ingredient of our system  -- i.e.,~the phase dependent correlated hopping has also been achieved experimentally \cite{Leinhard_PRX_2020}, potentially making such manipulation of the boundary modes not very far from an actual experimental realization. These results, therefore, are of particular relevance in context of the study of the effective low energy quantum mechanics of topological edge modes in one hand and their realization in ultracold atoms on the other. Devising concrete protocols for quantum gate operations in these low energy subspaces could be an interesting future direction which would allow such platforms to be used for quantum computation. 
Finally, we conclude by re-emphasizing that our study points out that particle statistics is an interesting handle to uncover the rich interplay of entanglement, topological order and role of symmetries in quantum few and many-body phenomena.

\acknowledgements
The authors acknowledge Gaurav K. Gupta and Vijay B. Shenoy for a previous collaboration. AA and SB acknowledge funding from Max Planck Partner Grant at ICTS.  SB acknowledges SERB-DST (India) for funding through project grant No. ECR/2017/000504. Numerical calculations were performed on clusters {\it boson} at ICTS. We acknowledge use of open-source QuSpin\cite{Weinberg_SP_2019} and ITensor\cite{ITensor} for ED and DMRG calculations respectively.  We acknowledge the International Centre for Theoretical Sciences (ICTS) for supporting the program- Geometric phase in Optics and Topological Matter (Code: ICTS/geomtop2020/01) and the Department of Atomic Energy, Government of India, under project no.12-R$\&$D-TFR-5.10-1100.

\begin{appendix}

\section{DMRG gap scaling}
\label{appenA}

Here in \Fig{fig:app1} we show the system size scaling of the energy gap $\Delta(1)$ (see \eqn{spg}) which quantifies the amount of energy required to add one particle about the ground state at half-filling for both periodic and open system. In \Fig{fig:figure4}(d) we had seen its behavior as a function of $\phi$ for a particular system size. In \Fig{fig:app1} we show the scaling of this gap as a function of $L$ showing that the single particle gap remains finite even in the thermodynamic limit. Under open boundary conditions, this excitation energy goes to zero when $t_1<t_2$ reflecting the existence of zero-energy boundary modes (see \Fig{fig:app1}(b)).

\begin{figure}
	\centering
	\includegraphics[width=1.0\linewidth]{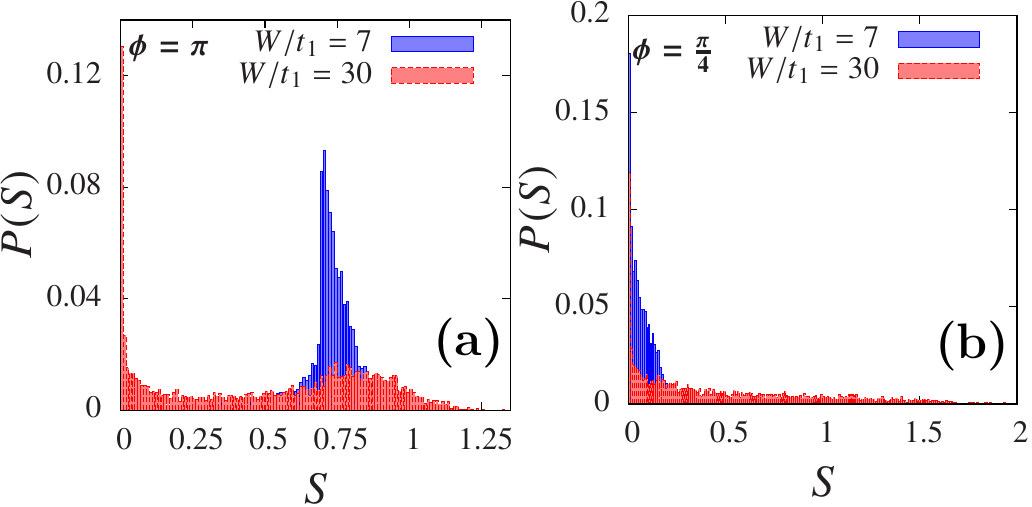}
	\caption{ {\bf Probability distribution of $S$:} The probability distribution of the mutual entanglement entropy $P(S)$ for the two-particle boundary mode ($e_1$) for different values of hopping disorder strength $W/t_1$ in (a) for $\phi=\pi$ and $W/t_1=7,30$ and in (b) for $\phi=\pi/4$ and $W/t_1=7,30$. The distribution at large $W/t_1$ is not a normal distribution. The sampling is done for 5000 independent disorder configurations. (ED, $t_1=0.1, t_2=1-t_1, \gamma=2,N=4L=40$, bin size=$0.01$)}
	\label{fig:app6}
\end{figure}

\section{Disorder Averaging} 
\label{appenB}

In \Fig{fig:figure3} we had shown the entanglement entropy of the boundary modes as a function of $\phi$ in presence of hopping disorder of strength $W$. Here, we systematically increase the value of $W$ and analyze the behavior of entanglement entropies in \Fig{fig:app2}. Given the presence of one particle on each chain; in \Fig{fig:app2}(a) we compare the energies of the near-zero energy modes and the gap to the bulk-excited states ($\equiv \delta$) at $\phi=0$. We find that the band-gap collapses near $W/t_1\sim 3$ for $t_1=0.1=1-t_2, \gamma=2$. We investigate the entanglement entropy as a function of $W/t_1$ for various values of  $\phi$ in \Fig{fig:app2}(c) and (d). While for low values of $W (\ll \delta)$  the fluctuations in the entanglement entropy remains small, but with large $W$ these fluctuations become significantly large. Investigating the probability distribution of $S$ (see \Fig{fig:app6}) we find that at large $W/t_1$, $S$ is far from any normal distribution consistent with studies of entanglement entropy in strongly disordered systems\cite{Laflorencie_PRB_2005, Pastur_JSP_2018}. It is useful to note entanglement entropy can be consistently defined only for non-degenerate eigenstates, therefore in sample averaging we discard states which are degenerate upto machine precision ($\sim 10^{-14}$). 

\end{appendix}

\bibliography{ref1DCorr}

\end{document}